\documentclass[a4paper,11pt]{article}
\pdfoutput=1 

\usepackage{jcappub} 

\usepackage[T1]{fontenc} 
\usepackage{placeins}
\usepackage[utf8]{inputenc}
\usepackage{amsmath,mleftright}
\usepackage{xparse}
\usepackage{caption}
\usepackage{subcaption}
\NewDocumentCommand{\evalat}{sO{\big}mm}{%
  \IfBooleanTF{#1}
   {\mleft. #3 \mright|_{#4}}
   {#3#2|_{#4}}%
}

\usepackage{graphicx}
\usepackage{dcolumn}
\usepackage{amssymb}
\usepackage{amsmath}
\usepackage{color}
\usepackage{verbatim}
\usepackage{multirow}
\usepackage{soul}
\usepackage{parskip}
\graphicspath{
{./LISA_window/}
{./figures_rough/}
{./figures_rough/sensitivity_curve/}
{./figures_rough/SNRs/}
{./figures_rough/parameter_relationships/mean_residual/}
{./figures_rough/parameter_relationships/}
{./figures/fit_comparison_plots/}
{./figures/relative_uncertainty_thermo_params/}
{./figures/relative_uncertainty_thermo_params/free_T/}
{./figures/SNRs/}
{./figures/varying_thermo_params/}
{./figures/relative_uncertainty_spectral_params/}
{./figures/suppression_factor/}
}

\def\al{\alpha} 
\def\be{\beta} 
\def\ga{\gamma}
\def\de{\delta}

\def\th{\theta}

\def\la{\lambda}
\def\rh{\rho}

\def\De{\Delta}

\def\Om{\Omega}

\def\pa{\partial}

\newcommand{\ben}{\begin{equation}}
\newcommand{\een}{\end{equation}}
\newcommand{\bea}{\begin{eqnarray}}
\newcommand{\eea}{\end{eqnarray}}
\newcommand{\ba}{\begin{array}}
\newcommand{\ea}{\end{array}}
\newcommand{\bit}{\begin{itemize}}
\newcommand{\eit}{\end{itemize}}

\newcommand{\A}{\mit{A}}
\newcommand{\E}{\mit{E}}
\newcommand{\T}{\mit{T}}

\newcommand{\Btilde}{\tilde{\beta}}

\newcommand{\cs}{c_\text{s}} 

\newcommand{\cj}{c_\text{J}}

\newcommand{\Dbin}{\bar{D}_{b}}
\newcommand{\DAn}{D^{\A}_n}
\newcommand{\DEn}{D^{\E}_n}
\newcommand{\DAbin}{\bar{D}^{\A}_b}
\newcommand{\DEbin}{\bar{D}^{\E}_b}

\newcommand{\fp}{f_\text{p}}
\newcommand{\fb}{f_\text{b}}
\newcommand{\fpt}{f_\text{p,0}}
\newcommand{\ft}{f_\text{t}} 
\newcommand{\fbin}{f_b} 

\newcommand{\HN}{H_\text{n}} 

\newcommand{\NA}{N_{\A}}
\newcommand{\NE}{N_{\E}}

\newcommand{\nbin}{n_{b}}
\newcommand{\Nbin}{N_\text{b}}

\newcommand{\OmGW}{\Omega_\text{gw}}
\newcommand{\OmGWSSM}{\Omega_\text{gw}^\text{ssm}}
\newcommand{\OmGWfit}{\Omega_\text{gw}^\text{fit}}

\newcommand{\OmGWscaled}{\tilde\Omega_\text{gw}}

\newcommand{\OmPeak}{\Omega_\text{p}}
\newcommand{\OmPeakt}{\Omega_\text{p,0}}

\newcommand{\Omeb}{\Omega_{\text{eb}}} 
\newcommand{\Omgb}{\Omega_{\text{gb}}} 

\newcommand{\Omt}{\Omega_{\text{t}}}
\newcommand{\OmInst}{\Omega_{\mathrm{ins}}}

\newcommand{\PspecGWhat}{\hat{\mathcal{P}}_{\text{gw}}}

\newcommand{\Pacc}{P_{\text{acc}}}
\newcommand{\Poms}{P_{\text{oms}}}

\newcommand{\Rstar}{R_*} 

\newcommand{\rb}{r_\text{b}} 

\newcommand{\SpecDenGW}{\tilde P_{\text{gw}}}

\newcommand{\SNR}{\rho}

\newcommand{\SA}{S_{\A}}

\newcommand{\Sn}{S_n} 
\newcommand{\Sbin}{S_{b}}

\newcommand{\TN}{T_\text{n}} 
\newcommand{\Tn}{T_\text{n}} 
\newcommand{\Tc}{T_\text{c}} 

\newcommand{\Tobs}{T_{\text{obs}}}

\newcommand{\vw}{v_\text{w}} 

\newcommand{\OmGWt}{\Omega_{\text{gw,0}}}
\newcommand{\Fgwt}{F_{\text{gw,0}}}

\newcommand{\zp}{z_\text{p}} 
\newcommand{\zb}{z_\text{b}} 

\definecolor{darkgreen}{rgb}{0.0,0.5,0.0}

\title{ Observational prospects for phase transitions at LISA: Fisher matrix analysis  }
\subheader{HIP-2021-17/TH}
\author[a]{Chloe Gowling,} 
\author[b,a]{Mark Hindmarsh.}

\affiliation[a]{Department of Physics and Astronomy, University of Sussex, BN1 9QH, Brighton, UK}
\affiliation[b]{Department of Physics and Helsinki Institute of Physics, PL 64, FI-00014 University of Helsinki,
Finland}

\emailAdd{c.gowling@sussex.ac.uk}
\emailAdd{mark.hindmarsh@helsinki.fi}

\abstract{A first order phase transition at the electroweak scale 
would lead to the production of gravitational waves that may be observable at upcoming space-based gravitational wave (GW) detectors such as LISA (Laser Interferometer Space Antenna). 
As the Standard Model has no phase transition, LISA can be used to search for new physics 
by searching for a stochastic gravitational wave background. 
In this work we investigate LISA's sensitivity to the 
thermodynamic parameters encoded in the stochastic background produced by a phase transition, 
using the sound shell model to characterise the gravitational wave power spectrum, and the Fisher matrix to estimate uncertainties. 
We explore a parameter space with transition strengths $\al < 0.5$ and phase boundary speeds $0.4 < \vw < 0.9$, 
for transitions nucleating at $\TN = 100$ GeV, with mean bubble spacings $0.1$ and $0.01$ of the Hubble length, 
and sound speed $c/\sqrt{3}$. 
We show that the power spectrum in the sound shell model can be well approximated by a 
four-parameter double broken power law, 
and find that the peak power and frequency can be measured to approximately 10\% accuracy for signal-to-noise ratios (SNRs) above 20.
Determinations of the underlying thermodynamic parameters are complicated by degeneracies, but in all cases 
the phase boundary speed will be the best constrained parameter. Turning to the principal components of the Fisher matrix, a signal-to-noise ratio above 20 
produces a relative uncertainty less than 3\% in the two highest-order principal components, 
indicating good prospects for combinations of parameters.  
The highest-order principal component is dominated by the wall speed. These estimates of parameter sensitivity provide a preliminary accuracy target for theoretical calculations of thermodynamic parameters.
}

\begin{document}
\maketitle
\flushbottom

\setlength{\parskip}{6pt plus 1pt}

\section{Introduction}

\label{sec:intro}
With the initial LIGO detection of a black hole merger \cite{Abbott:2016blz}, the multi-messenger observation of a merging neutron star binary \cite{GBM:2017lvd} and the recent observation of an intermediate mass black hole merger \cite{Abbott:2020khf} we are beginning to realise the discovery potential of gravitational waves (GWs).  LIGO and other ground-based detectors are optimised for stellar-origin black holes and sensitive to the $10$ Hz -$10$ kHz frequency band. The exploration of signals from the merger of the much larger black holes at the centre of galaxies 
will take place at future space-based GW observatories, where longer arm lengths open up the $10^{-4}$ Hz to $10^{-1}$ Hz range of the GW spectrum. Such experiments include the ESA-NASA mission LISA \cite{Audley:2017drz}, 
Taiji \cite{Guo:2018npi} and TianQin \cite{Mei:2020lrl}, all aiming for launch in the mid-2030s. LISA and TianQin have both launched test satellites,  the final LISA pathfinder mission results  \cite{Anderson:2018nqj} and the initial results from TianQin-1 \cite{Luo:2020bls} are promising.

As well as massive black hole mergers, expected astrophysical sources in the millihertz band include galactic binaries \cite{Postnov:2014tza}, extreme mass ratio binaries \cite{AmaroSeoane:2007aw} and precursors for stellar origin black hole mergers \cite{Sesana:2016ljz}.  Cosmological sources could include stochastic gravitational wave backgrounds (SGWBs) from inflation, cosmic strings and cosmological phase transitions \cite{Caprini:2018mtu}.  

In this paper we focus on a SGWB from cosmological phase transitions, specifically those from around 10 picoseconds after the Big Bang, when it is expected the electroweak symmetry broke.  In the Standard Model this process occurs via crossover \cite{Kajantie:1996mn,Kajantie:1996qd}, and the GW signal is expected to be negligible at observable frequencies \cite{Ghiglieri:2020mhm}. However, there are numerous extensions to the Standard Model in which a first order phase transition is possible. A  review of possible extensions can be found in Ref.~\cite{Caprini:2019egz}.
 
In such theories, below the critical temperature, bubbles of the stable phase spontaneously nucleate in the surrounding metastable phase. These bubbles expand, collide and merge until only the stable phase remains, leaving behind a characteristic spectrum of sound waves, which are a persistent source of GWs \cite{Hindmarsh:2013xza,Hindmarsh:2015qta,Hindmarsh:2017gnf}.  The collision of bubble walls \cite{Kosowsky:1991ua,Cutting:2020nla,Lewicki:2020jiv,Lewicki:2020azd} and turbulent flows 
\cite{Kosowsky:2001xp,Gogoberidze:2007an,Caprini:2009yp,Pol:2019yex} also generate GWs. 
Here, we consider only  the contribution from sound waves as they are currently expected to be the dominant source 
over a wide range of parameters \cite{Caprini:2019egz}. 

If the critical temperature is in the range 100 -- 1000 GeV, the peak frequency  of the GW power spectrum can be in the millihertz range, and potentially detectable at a space-based observatory. This means that the discovery potential of GW observations includes fundamental physics beyond the Standard Model. The new physics may include a mechanism for baryogenesis \cite{Kuzmin:1985mm}, a strong motivation for 
considering Standard Model extensions with a first order phase transition.
For a recent introduction to baryogenesis see Ref.~\cite{Cline:2018fuq}, 
and to phase transitions in the early Universe see Refs.~\cite{Mazumdar:2018dfl,Hindmarsh:2020hop}. 
For a review of the prospects for probing physics beyond the Standard Model, see Ref.~\cite{Caprini:2019egz}.

Numerical simulations for the acoustic contribution to the GW signature from a first order phase transition \cite{Hindmarsh:2013xza,Hindmarsh:2015qta,Hindmarsh:2017gnf} have shown that the sound waves generated by the expanding bubble determine the GW power spectrum. 
The simulations motivate a simple broken power-law model for the sound wave contribution used by the LISA cosmology working group \cite{Caprini:2019egz}. The uncertainties that arise from  various levels of approximations have been explored in \cite{Guo:2021qcq}. 

Currently the most sophisticated model for computing the GW  power spectrum from sound waves 
is the sound shell model (SSM) \cite{Hindmarsh:2016lnk,Hindmarsh:2019phv}. 
The SSM shows how the GW power spectrum can be computed from the velocity power spectrum of the fluid, which in turn is dependent on a few key thermodynamic parameters that can be calculated from the underlying theory. 
These key parameters effect the overall amplitude, frequency scale and the detailed shape of the power spectra. 
In its simplest form, there are four thermodynamic parameters: the bubble nucleation temperature, 
the transition rate, the transition strength, and the bubble wall speed. All are in principle computable from an 
underlying theory, making them the interface between observation and theory. 
At the moment there are significant uncertainties in these calculations  \cite{Croon:2020cgk}. 
Our work can be used to set targets for future developments of theoretical methods.

The SSM predicts two important frequency scales in the power spectrum, and a double broken power law has been proposed as an analytic fit \cite{Hindmarsh:2019phv}. The functional form depends on the peak power, peak frequency, the ratio of the frequencies of the two breaks 
and the slope between the two breaks. We call these the spectral parameters, and distinguish them from the thermodynamic parameters discussed above. We show that the double broken power law form is much closer to the SSM prediction than 
the single broken power law fit given in \cite{Caprini:2019egz}. 

In this work we use the Fisher matrix \cite{Fisher:1922saa} to explore LISA's ability to extract parameters 
that describe a SGWB from a first order phase transition, 	also examining the effect of 
the expected foregrounds from galactic and extragalactic compact binaries. The Fisher matrix is known to overestimate uncertainties, especially when there are degeneracies amongst parameters, as is thought to be the case with the thermodynamic parameters. Despite this, we can expect the Fisher matrix will give an insight into parameter sensitivity and provide a better understanding of the degeneracies themselves.

We calculate the relative uncertainty both of the spectral parameters and the thermodynamic parameters as described above, 
with and without foregrounds, over a range of fiducial models.
We focus on LISA but the methods could be easily adapted to other missions by altering the noise model. This complements general power law searches in mock LISA data  \cite{Adams:2013qma, Boileau:2021sni, Boileau:2020rpg} Fisher matrix analysis for a single broken power law with LISA, DECIGO and BBO mock data  \cite{Hashino:2018wee}, searches for cosmological phase transition SGWB in LIGO and NANOGrav data \cite{Romero:2021kby,Arzoumanian:2021teu}, and methods for general SGWB searches where the search is agnostic about the spectral shape of the GW background \cite{Pieroni:2020rob,Flauger:2020qyi}.

For our fiducial models 
we focus on a thermodynamic parameter space motivated by an electroweak-scale transition, by relevance for observation, 
and also by the reliability of predictions. 
The electroweak scale motivates the choice of nucleation temperature $\TN=100$ GeV. 
Relevance for observation motivates examining supercooled transitions with 
mean bubble spacing to Hubble length ratio $r_* = 10^{-1}$ and $10^{-2}$, as much smaller values would render the signal too weak.
The reliability of the sound shell model predictions can be tested against numerical simulations \cite{Cutting:2019zws} 
in the range of wall speeds $0.24 < \vw < 0.92$ and with transition strength parameter $\al < 0.5$. 
We study the range $0.4 < \vw < 0.9$, as lower wall speeds will also probably not be observable at LISA. 

This parameter space produces signals with gravitational wave density fraction today up to $\OmPeakt \sim 10^{-10}$ and 
peak frequencies $\fpt$ in the range $10^{-2}$ mHz to $5$ mHz, which can produce SNRs well over 100. 
A transition with $r_* = 0.1$ should produce an observable signal over most of the 
parameter space.

Of the spectral parameters, LISA will be most sensitive to the peak power and peak frequency, reaching 
approximately 10\% uncertainty in the peak power and frequency for signal-to-noise ratio (SNR) above 20. 
For $r_* = 0.1$, SNR 20 can be reached over most of the range $0.5 < \vw < 0.8$ and $\al > 0.2$.  
For $r_*= 0.01$, stronger transitions are required to reach the same SNR.

Of the thermodynamic parameters, there is greatest sensitivity to the wall speed.  
At SNR $=20$ the relative uncertainty in the wall speed is $10\%$ in some regions of the thermodynamic parameter space, 
but the sensitivity to the other parameters is reduced by degeneracies. 
Examining the principal components, 
one finds an uncertainty of $3\%$ or better for the two highest-order components, at SNR $=20$. 
Hence there are good prospects for combinations of parameters. 
The best-determined principal component is dominated by the wall speed. The second-best has $\al$ as the most important contribution, but other parameters also contribute.

If a parameter combination could be predicted in the light of other data, the prospects for estimating the other parameters would be much better. 
As a simple example, we consider a case where the nucleation temperature is known, 
for a transition in which the  mean bubble spacing parameter is  $r_* = 0.1$.
Here, the phase transition strength and the mean bubble spacing can be constrained to $10\%$ and $30\%$ respectively. 
If the galactic binary foreground can be removed, 
the uncertainty in the phase transition strength can be as low as 10\% for transitions with $\al \simeq 0.1$. 

The paper is organised as follows. In Sec.~\ref{sec:cosmo} we review the production of GWs from a first order phase transition in the early universe, how they relate to the underlying thermodynamic parameters, and introduce the SSM \cite{Hindmarsh:2016lnk}. The setup we consider for LISA and the noise model  is outlined in Sec.~\ref{sec:noise}. In Sec.~\ref{sec:fm} we describe our method for calculating the Fisher matrix, relative uncertainties and principal components. The relative uncertainties in the spectral and thermodynamic parameters are presented in Sec.~\ref{sec:results}. The discussion of the results are given in Sec.~\ref{sec:Discussion}.

In this work we set $c= 1$ and $k_{\text{B}} =1$, unless otherwise specified. 

\section{Gravitational waves from a first order cosmological phase transition}
\label{sec:cosmo}
\subsection{Cosmological phase transitions }
As the universe expanded and cooled, significant changes of in the equation of state must have occurred at temperatures of around 100 GeV, when elementary particle rest masses were generated, and at 
100 MeV, when quarks and gluons became confined into hadrons.  

It is known that both of these changes happened via a smooth cross-over in the Standard Model  
\cite{Borsanyi:2016ksw}, \cite{Kajantie:1996qd,Kajantie:1996mn}, but in extensions of the Standard Model 
first order electroweak-scale transitions are common (see Ref.~\cite{Caprini:2019egz} for a survey). Such phase transitions are often associated with a change in the symmetry of the 
plasma, accompanied by a change in the value of an order parameter, which in the 
case of the electroweak transition is the magnitude of the Higgs field. 

In a symmetry-breaking phase transition such as the electroweak transition, one often refers to the high-temperature phase as the ``symmetric'' phase and the low-temperature phase as the ``broken'' phase. 
 
In a first order phase transition, at a critical temperature  $T_\text{c}$ there are 
two degenerate minima of the free energy separated by a barrier.  
As the temperature cools below $\Tc$, the broken phase becomes lower in free energy, and the 
system can move to it via localised thermal or quantum fluctuations. 
This leads to bubbles of broken phase nucleating within the symmetric region. These bubbles expand due to the pressure difference between the interior and exterior, inevitably present as the pressure is minus the free energy density. 
The bubbles collide and merge until only the broken phase remains. 
Some of the latent heat of the transition is converted into kinetic energy of the cosmological fluid surrounding the bubbles, which is a source of shear stress, leading to the production of gravitational waves. 

Now we introduce the key thermodynamic parameters that determine the gravitational wave signature from a first order phase transition (see e.g.~\cite{Hindmarsh:2020hop}). The first of these is the nucleation temperature $\TN$, which we define as the peak of the globally-averaged bubble nucleation rate. The Hubble rate at the nucleation temperature sets the frequency scale of the GW power spectrum. 

The second one is the nucleation rate parameter $\beta$, which is often given as a ratio with $\HN$, the Hubble parameter at $\TN$
\ben{\label{Eq:B/H}}
   \Btilde=\frac{\beta}{\HN} \sim  \frac{\vw}{\HN R_*},
\een
where $\vw$ is the speed of the expanding bubble wall and $R_*$ is the mean bubble spacing. From this we see that $\tilde{\beta}$ controls $R_*$, which  in turn sets the characteristic wavelength of the gravitational radiation. The constant of proportionality is $(8\pi)^{1/3}$ \cite{Enqvist:1991xw} for detonations, but 
for deflagrations it is also dependent on $\al$ and $\vw$, as the nucleation rate is reduced by 
the reheating of the fluid in front of the bubble wall. In view of this uncertainty, it is more convenient to work in terms of $R_*$, and more precisely 
the Hubble-scaled mean bubble spacing
\ben\label{Eq:rstar}
r_* = \HN R_*.
\een
Note that $\be^{-1}$ is the time taken for a bubble wall to move a distance $R_*$, and therefore has 
an interpretation as the duration of the phase transition. 

Another key parameter in  the generation of GWs is the phase transition strength parameter $\al$
  \ben{\label{Eq:alpha}}
 \alpha=\left.\frac{4}{3} \frac{\Delta \theta}{w_{\mathrm{s}}}\right|_{T=T_{\mathrm{n}}}
 \een
where  $w_\text{s} $ is  the enthalpy of the fluid in the symmetric phase, and  $\Delta \theta=\theta_{\mathrm{s}}-\theta_{\mathrm{b}}$, where $\th$ is a quarter of the trace of the energy-momentum tensor, 
and subscripts s and b denote symmetric and broken phases. 
The trace difference is the energy available to be converted to shear stress energy and thus GW power. A stronger transition means more energy is converted to shear stress energy and a larger overall amplitude for the GW signal. 

The fourth parameter is the wall speed, $\vw$, which (with $\al$) determines the motion of the surrounding plasma induced by the passing bubble wall. Wall speeds are split into three categories relative to the speed of sound $\cs$. Deflagrations occur when $\vw<\cs$, where the surrounding fluid is pushed in front of the expanding phase transition wall. When $\vw$ is greater than a certain critical speed $\cj$, the Jouguet speed, 
the motion in the plasma is entirely behind the bubble wall, and the fluid configuration is called a detonation.
The Jouguet speed is given by 
\ben\label{Eq:Jouguet}
\cj = c_\text{s} \frac{\left(1 + \sqrt{\al(2 + 3\al) }\right)}{\left(1 + \al\right)}
\een
If the wall speed is between the sound speed and the Jouguet speed, 
the velocity profile is a mix between deflagrations and detonations, non-zero both in front and behind the bubble wall.  
These supersonic deflagrations \cite{KurkiSuonio:1995pp}, sometimes called hybrids \cite{Espinosa:2010hh}, 
are very finely tuned, and it is not clear that they exist in a real fluid. 

The sound speeds in the two phases are also potentially important parameters \cite{Giese:2020rtr,Giese:2020znk}. 
To simplify this first analysis, we will take them both to be the ultrarelativistic value $\cs = 1/\sqrt{3}$, 
to focus on LISA's sensitivity to the four parameters ($\TN$,$\al$,$r_*$,$\vw$). 
\subsection{Gravitational waves from a first order phase transition}

In a first order transition driven by thermal fluctuations, sound waves created by the expanding bubbles 
are the dominant source of gravitational waves \cite{Hindmarsh:2013xza,Hindmarsh:2015qta,Hindmarsh:2017gnf}.

Approximate fits to the numerical power spectra are available in \cite{Hindmarsh:2017gnf}. They have a fixed broken power law shape, with peak intensity and frequency depending on 
the four thermodynamic parameters in an easily computable way \cite{Caprini:2019egz}. 
The peak intensity depends on $\al$, $r_*$ and $\vw$, while the peak frequency 
depends on $\TN$ and $r_*$.  
It is clear that there are likely to be degeneracies in the power spectrum with respect to 
the thermodynamic parameters, which would intrinsically limit 
LISA's ability to measure them individually. 

However, the simulations make it clear that the shape of the GW power spectrum also 
depends on wall speed and transition strength, 
and such dependence is found in a more sophisticated theoretical framework, 
the sound shell model (SSM) \cite{Hindmarsh:2016lnk,Hindmarsh:2019phv}.
We therefore use the sound shell model to model the GW power spectrum from phase transitions,
and investigate LISA's constraining power on its parameters. 
While the sound shell model has not been tested in detail against a wide range of numerical simulations, 
it can act as guidance for data analysis techniques aimed at 
extracting phase transition parameters from phase transitions.  

To characterise how the energy density in GWs is distributed over frequencies today we introduce the 
gravitational wave power spectrum \cite{Allen:1997ad}
 \ben{\label{Eq:GW_en_dens}}
\OmGWt (f) \equiv \frac{1}{\rho_{\text{c},0}}\frac{d  \rho_{\text{gw},0}}{d \ln f},
\een
where $f$ is frequency and $d\rho_{\text{gw}}$ is the gravitational wave energy density within a frequency interval $df$. 
The critical density is $\rho_{\text{c}} = {3H^2}/{8\pi G}$, where $H$ is the Hubble rate, $G$ is the gravitational constant and $c$ is the speed of light. Quantities evaluated at the present day are given the subscript 0. For the Hubble constant 
$H_0$  we take the central value measured by the Planck satellite $H_0= 67.4 \, \text{km s}^{-1}\text{Mpc}^{-1}$ as given in \cite{Aghanim:2018eyx}.

 The general form of the gravitational wave power spectrum from a first order phase transition 
 is
 \ben\label{Eq:Omgw_ssm}
 \OmGW(z) = 3K^{2}(\vw,\al)\left(\HN \tau_{\mathrm{v}}\right)\left(\HN R_*\right) \frac{z^{3}}{2 \pi^{2}} \SpecDenGW\left(z\right),
 \een
 where $R_*$ is the mean bubble spacing, 
 $z = k R_*$, $k$ is comoving wavenumber and $K(\vw,\al)$ 
 is the fraction of the total energy converted into kinetic energy of the fluid. 
 The Hubble rate at nucleation is $\HN$, $\tau_v $ is the lifetime of the shear stress source, 
the factor $R_*$ appears as an estimate of the source coherence time and $\SpecDenGW \left(z\right)$ is the dimensionless spectral density. 
Eq.~(\ref{Eq:Omgw_ssm}) can be regarded as the definition of $\SpecDenGW$. 
 Its integral (denoted $\OmGWscaled$ in Refs.~\cite{Hindmarsh:2015qta,Hindmarsh:2017gnf,Hindmarsh:2019phv}) 
depends only weakly  on the thermodynamic parameters, taking values of order $10^{-2}$. 

As the notation of Eq.~(\ref{Eq:Omgw_ssm}) suggests, the important parametric dependences of the total power are 
through the kinetic energy fraction, the source lifetime, and the source coherence time.
The kinetic energy fraction depends only on the transition strength $\alpha$ and the wall speed $v_w$.
The lifetime of the GW source $\tau_v$ is the shorter of the two timescales, 
the Hubble time $\HN^{-1}$ and the fluid flow lifetime  $\tau_v$, 
which is estimated as $R_* / \sqrt{K}$, the timescale for 
non-linearities to become important.
Denoting the ratio of the two timescales by $x = \HN R_* / \sqrt{K} $, 
we approximate the Hubble-scaled source lifetime as \cite{Guo:2020grp}
 \ben{\label{Eq:Hntv}}
\HN \tau_v \simeq \left(1 -  \frac{1}{\sqrt{1 + 2x}} \right).
 \een
From this we see that even if the flow persists over many Hubble times it does not 
continue to contribute to the GW power spectrum. 
For future convenience we will combine the factors of the source lifetime and source coherence time into one, 
\ben\label{Eq:scaling_factors_GW}
J = H_n R_* H_n\tau_v  = r_* \left(1 -  \frac{1}{\sqrt{1 + 2x}} \right).
\een

The sound shell model \cite{Hindmarsh:2016lnk,Hindmarsh:2019phv} predicts the gravitational wave power spectrum as 
a numerical function of a given set of thermodynamic parameters ($\TN$, $\alpha$, $r_* $, $\vw$) and 
scaled wavenumbers $z$.  We denote this prediction $\OmGWSSM(z)$.  
The shape of the power spectrum has significant dependencies on $\vw$ and $\alpha$.
 
Recent 3d-hydro simulations for $\al$ up to $\mathcal{O}$(1) (strong transitions) found that as transition strength increases the efficiency of fluid kinetic energy production becomes less than previously expected \cite{Cutting:2019zws}. For deflagrations this is thought to be due to reheating which occurs in front of the expanding bubbles, which leads to a reduction in pressure difference, and a slowing of the bubble wall. The reduction in kinetic energy production leads to a suppression in gravitational waver power, which we approximate by a factor $\Sigma(\vw,\al)$. The estimation of this suppression factor from the numerical simulations is described in Appendix \ref{sec:suppression_factor}. 

The gravitational wave power spectrum at dimensionless comoving wavenumber $z$ just after the transition, and before 
any further entropy production, is then
\ben\label{Eq:SSM_suppressed}
\Om_{\text{gw}}(z) = \OmGWSSM(z)\Sigma(\vw,\al),
\een
where $ \OmGWSSM(z)$ is the sound shell model prediction.

Today the power spectrum at physical frequency $f$ is 
\ben{\label{Eq:Omgw0_sup}}
    \OmGWt(f) =\Fgwt \Om_{\text{gw}}(z(f)),
\een
where 
\ben{\label{Eq:Fgw0_def}}
\Fgwt=\Omega_{\gamma, 0}\left(\frac{g_{s 0}}{g_{s *}}\right)^{\frac{4}{9}} \frac{g_{*}}{g_{0}} = (3.57 \pm 0.05) \times 10^{-5} {\bigg( \frac{100}{g_*}\bigg)}^{\frac{1}{3}} ,
\een
is the power attenuation following the end of the radiation era. 
Here, $\Omega_{\gamma, 0}$ is the photon energy density parameter today, $g_{s }$ denotes entropic degrees of freedom and $g$ describes the pressure degrees of freedom. 
In both cases the subscripts $0$ and  $*$ refer to their value today and the value at the time the GWs were produced respectively. 
We evaluate $\Fgwt$ with the values given in \cite{Caprini:2019egz}, and use a reference value $g_* = 100 $. 

We convert from dimensionless wavenumber $z$ to frequency today by taking into account redshift
\ben
\label{e:fzrstar}
f =\frac{z }{r_*} f_{*,0}
\een
where \cite{Caprini:2019egz}
\ben {\label{Eq:f0} }
f_{*,0}=  2.6 \times 10^{-6} \,\textrm{Hz} \left(\frac{\TN}{100\,\textrm{GeV}}\right)\left(\frac{g_*}{100}\right)^{\frac{1}{6}}
\een
is the Hubble rate at the phase transition redshifted to today. 
We assume the phase transition takes place well within one Hubble time so all frequencies throughout the transition have the same redshift. 

\begin{figure}[b!]
  \begin{subfigure}{.5\textwidth}
    \centering
    \includegraphics[width=\linewidth]{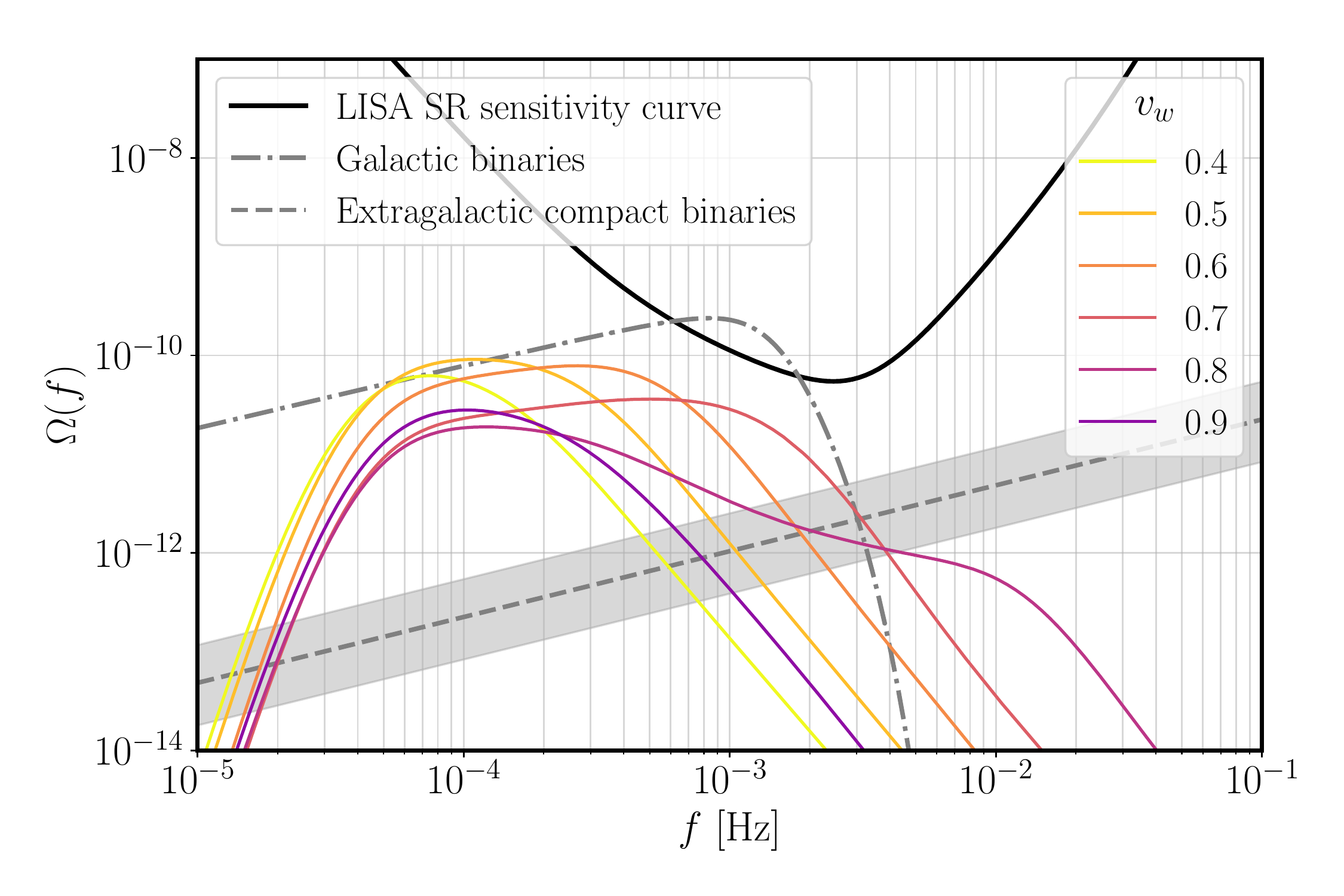}
    \caption{\label{Fig: vary_vw} Fixed: $\al =0.2, \quad r_* =0.1, \quad \TN=100 $ GeV. }
  \end{subfigure}
  \hfill
  \begin{subfigure}{.5\textwidth}
    \centering
    \includegraphics[width=\linewidth]{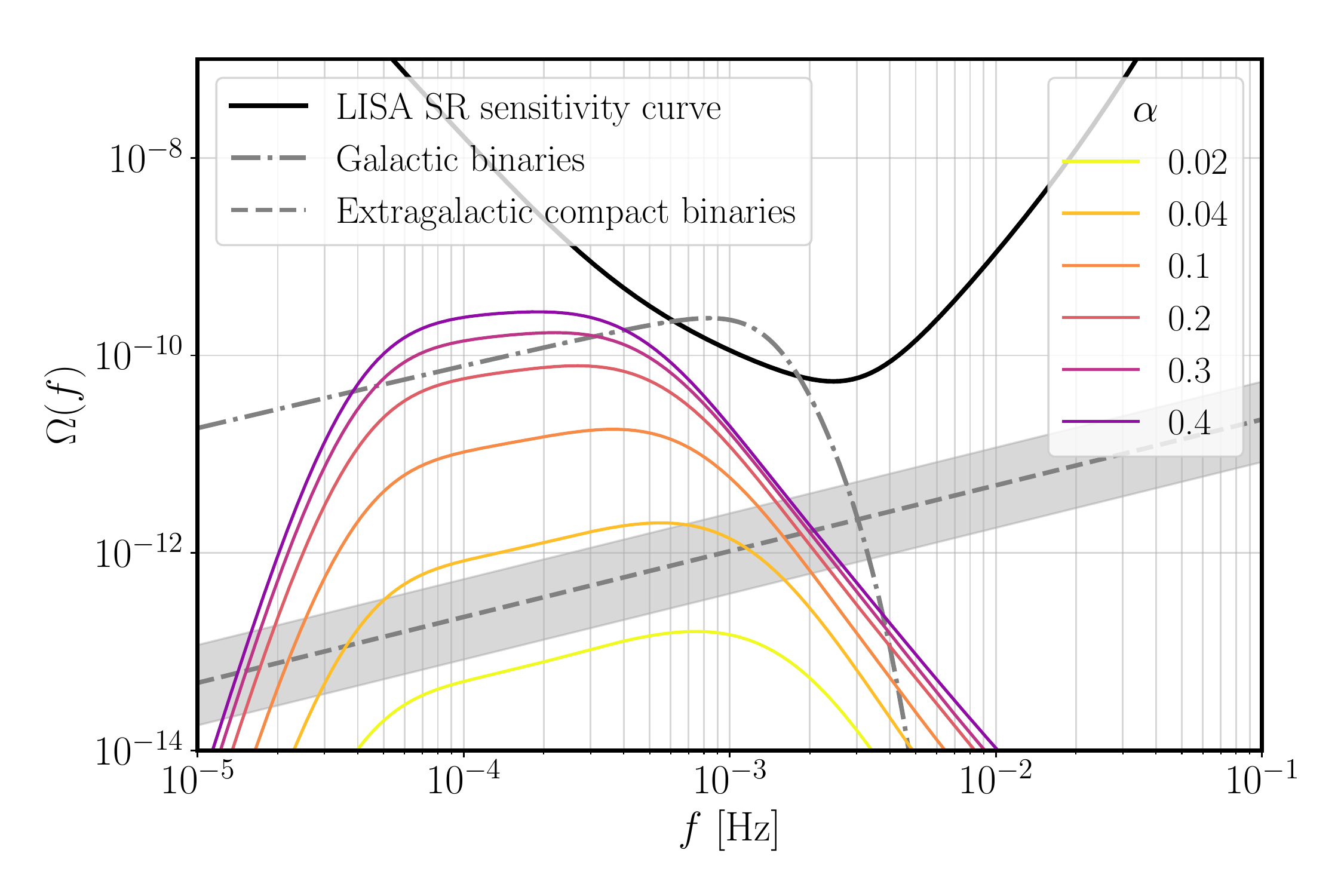}
    \caption{\label{Fig: vary_al} Fixed: $\vw = 0.6, \quad r_* =0.1, \quad \TN=100$ GeV. }
  \end{subfigure}

  \medskip

  \begin{subfigure}{.5\textwidth}
    \centering
    \includegraphics[width=\linewidth]{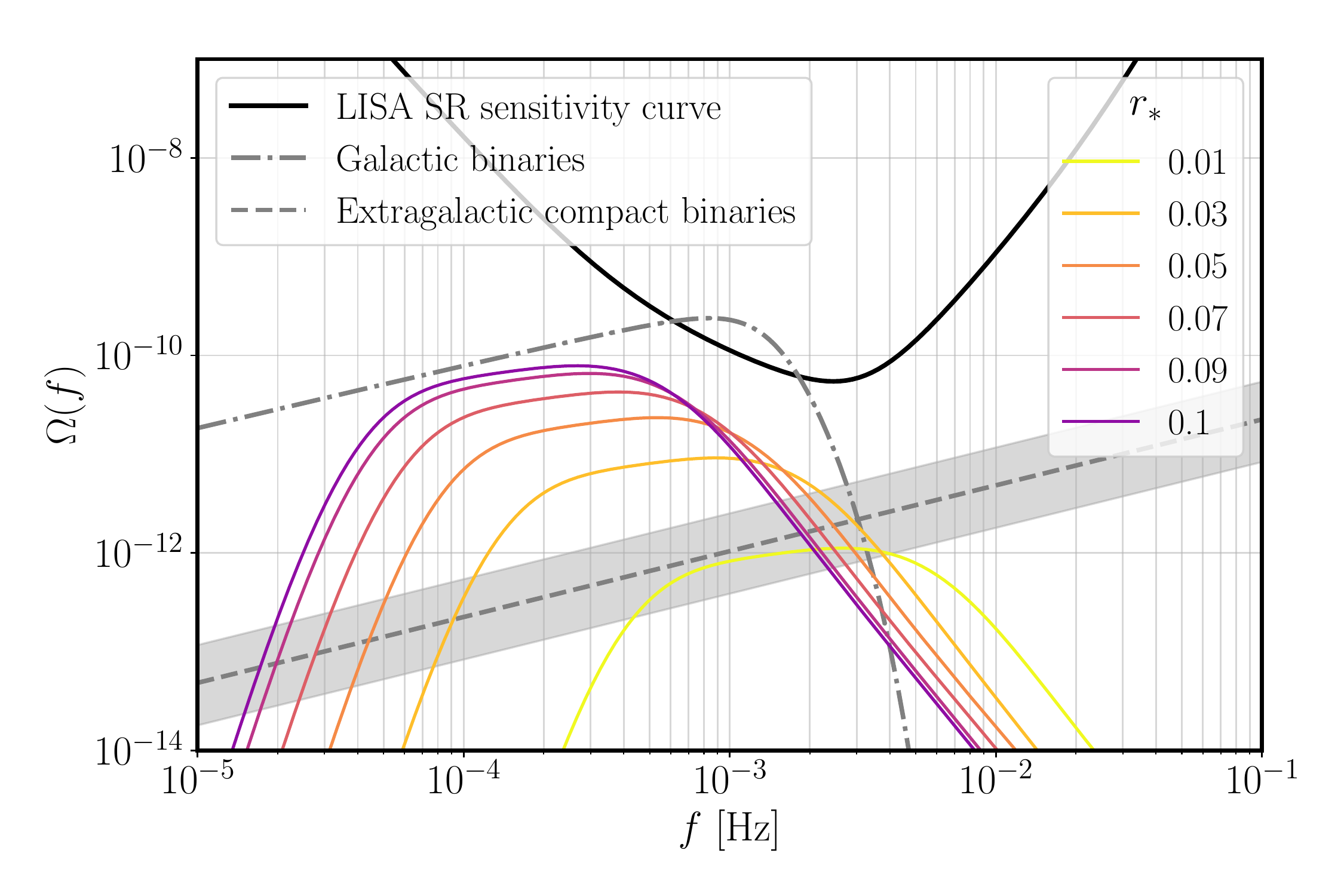}
    \caption{\label{Fig: vary_rs}Fixed:  $\vw = 0.6, \quad \al =0.2, \quad \TN=100 $ GeV. }
  \end{subfigure}
  \hfill
  \begin{subfigure}{.5\textwidth}
    \centering
    \includegraphics[width=\linewidth]{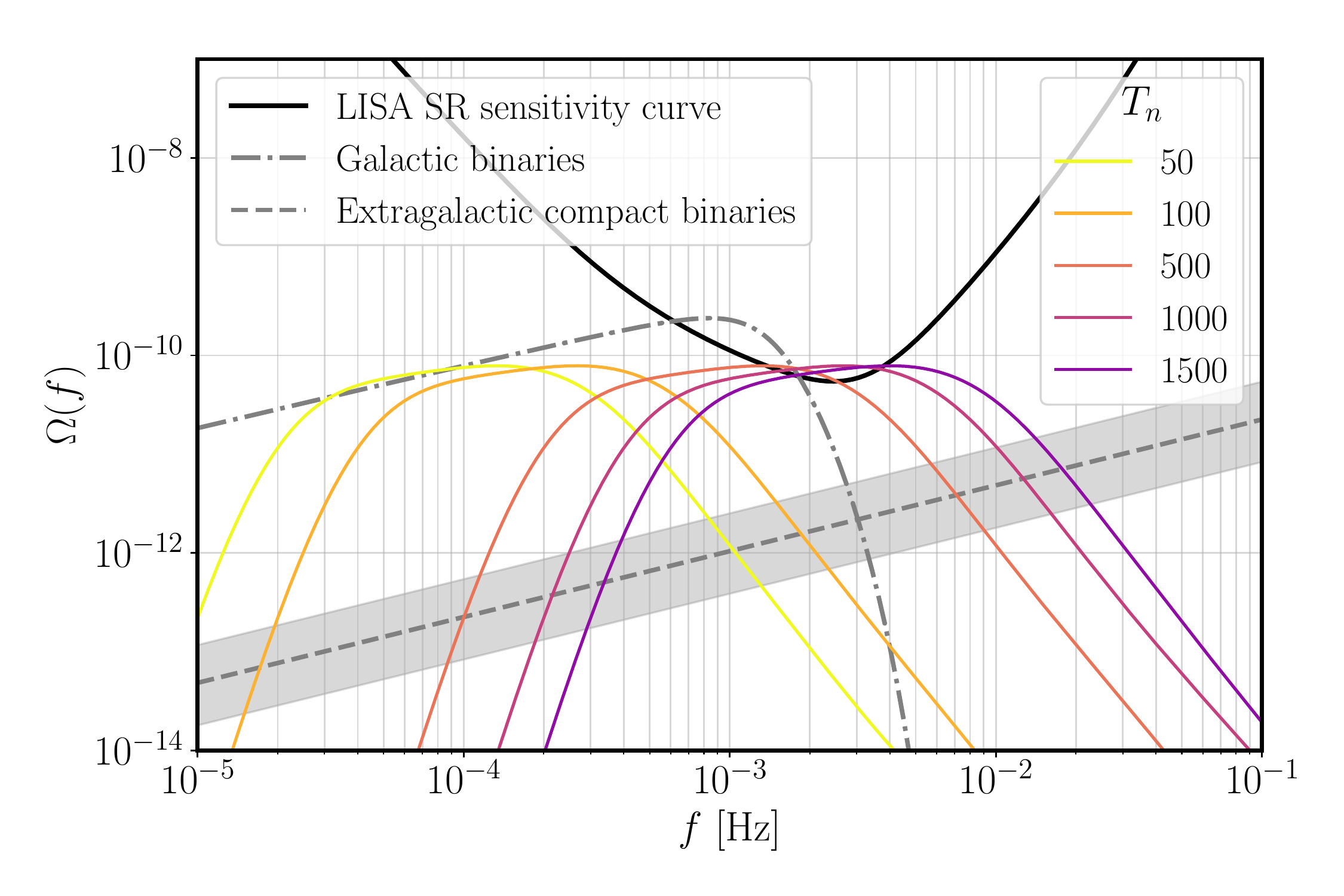}
    \caption{\label{Fig: vary_Tn}Fixed: $\vw = 0.6, \quad \al =0.2, \quad r_* =0.1$. }
  \end{subfigure}
   \caption{{\label{Fig:parameter_vary}} Gravitational wave power spectra for a first order phase transition calculated using the sound shell model, Eq.(\ref{Eq:Omgw0_sup}). In each panel we vary one of the thermodynamic parameters $\vw$ (wall speed), $\al$ (phase transition strength), $r_*$ (Hubble-scaled bubble spacing) and $\TN$ (nucleation temperature). Shown also in solid black is the LISA instrument noise given by the science requirements (SR) document  sensitivity curve (Eq.~(\ref{Eq:OmInst}), \cite{LISA_SR_doc}). The dashed line shows the predicted foreground from extragalactic binaries, Eq.~(\ref{Eq:Om_LV}), along with a grey uncertainty band. The dash-dotted line shows the estimated foreground from unresolved galactic binaries, Eq.~(\ref{Eq:Om_gb}).
Signal-to-noise ratios for $\TN = 100$ GeV and $r_* = 0.1, 0.01$ are given in Fig.~\ref{fig:SNRs}. 
}
\end{figure}

We compute the scale-free gravitational wave spectral density
\ben{\label{Eq:power_gw_scaled_output}}
\PspecGWhat (z)= 3K^{2} \frac{z^{3}}{2 \pi^{2}} \SpecDenGW \left(z\right)  , 
\een
using the PTtools python module \cite{Hindmarsh:2019phv}.  
 In this work we evaluated the power spectra at $100 $ logarithmic spaced $z$ values between $1$ and $1000$.
 The number of points used in the fluid shell profiles was $70000$, with $7000$ wavevectors used in in the velocity convolution integrations.  The bubble lifetime distribution, taken to be exponential, was integrated with $200$ linearly spaced values between $0$ and $20\beta^{-1}$. 
The high wavenumber resolution was used to ensure the integration over the velocity power spectrum converges. 
As mentioned in the introduction, we explore the prospects for estimation in the parameter space $0.4 < \vw < 0.9$,  $\alpha < 0.5$, $r_* = 0.01,0.1$ and $\TN = 100$ GeV.

We show, using this framework to calculate the GW power spectra,  how varying the thermodynamic parameters effects the shape, frequency scales and amplitudes of the power spectrum in Fig.~\ref{Fig:parameter_vary}.

From Fig.~\ref{Fig: vary_vw} we see the wall speed $\vw$ has a strong effect on the shape of the power spectrum, 
especially between the sound speed and the Jouguet speed. 
At low $\vw$ the power spectrum is narrow and as $\vw$ approaches the speed of sound the peak broadens, due to the 
narrowing of the sound shell around expanding bubbles. 
Once $\vw>\cs$ the peak begins to narrow again. As $\vw$ increases we also see a decrease in overall amplitude, 
because the efficiency of converting 
latent heat into fluid motion depends on $\vw$. 

As the strength of the phase transition, $\al$, increases so does the overall amplitude of the GW power spectrum, as more kinetic energy is deposited into the plasma (see Fig.~\ref{Fig: vary_al}). In Fig.~\ref{Fig: vary_rs} we note that $r_*$ contributes both to the frequency scale and overall amplitude of the power spectrum. In Fig.~\ref{Fig: vary_Tn} we see that the nucleation temperature $\TN$ affects only the frequency scale see Eq.(\ref{Eq:f0}).

We note that there is more structure in these power spectra than can be captured by a broken power law, motivating 
an improved approximation in the next section. 
The precise functional dependence on the thermodynamic parameters is likely to change as our understanding improves, 
but our analysis can be easily adapted to include future developments. 
We believe that the double broken power law form is 
likely to remain adequate.

\subsection{Double broken power law }
\label{ssec:cosmo_dbl_brkn}
The full calculation in the SSM can be computationally intensive when one is calculating many power spectra over a large parameter space. This motivates the use of an analytic fit that that can be used for rapid evaluation. The LISA Cosmology Working Group 
put forward a single broken power law to describe the acoustic contribution to the GW power spectrum, with two parameters, the peak amplitude $\OmPeak$ and the peak frequency $\fp$, whose scale is set by the bubble spacing $R_*$  \cite{Caprini:2019egz}.

In the SSM there are in fact two characteristic length scales, $R_*$ and the width of the sound shell $\De\Rstar \sim |\vw - \cs|/\be$, which indicate a double broken power law may be a good fit for the power spectrum \cite{Hindmarsh:2019phv}.  A general form for such a double broken power law can be defined by four spectral  parameters $(\OmPeak, \fp, \rb, b)$, with the power spectrum taking the form 
\ben{\label{Eq:omgw_dbl_brkn}}
    \OmGWfit =\Fgwt\OmPeak M(s,\rb, b)
\een
where $\OmPeak$ is the peak power of the power spectrum, $s = f/\fp$, $\fp$ is the frequency corresponding to $\OmPeak$ and  $\rb =  f_{\text{b}} /\fp$ describes the ratio between the two breaks in the spectrum. The parameter $b$ defines the spectral slope between the two breaks. The spectral shape $M(s,\rb, b)$ is a double broken power law with a spectral slope $9$ at low frequencies and $-4$ at high frequencies. 
\ben{\label{Eq: M double_break}}
    M ( s, \rb , b ) = s^ { 9 } {\left( \frac { 1 + \rb^4 } { \rb^4 + s^4}\right)}^{(9 -b)/4}  \left( \frac { b +4 } { b + 4 - m + m s ^ { 2 } } \right) ^ { (b +4) / 2 } .
\een
In this function, $m$ has been chosen to ensure that  for $\rb<1$ the peak occurs at $s=1$  and $M(1,\rb,b) = 1$, giving 
\ben{\label{Eq: m}}
    m = \left( 9 {\rb}^4+ b\right) / \left( {\rb}^4 +1 \right).
\een

Ultimately, we want to connect these spectral parameters quantitatively with the thermodynamic parameters in order to understand the underlying theory, however these relationships are not straightforward. An outline of how the spectral parameters depend on the thermodynamic parameters is as follows 
\ben {\label{Eq:spectral_parameters_thermo}}
\begin{aligned}
\OmPeakt &= \Fgwt J\left(r_{*}, K\left(\alpha, \vw\right)\right) \hat{\Omega}_{\text{p}}\left(\alpha, v_{w}\right) \Sigma_\text{ssm}(\vw,\al)\\
\fpt &= f_{*,0}(\TN)z_{\text{p}}\left(\alpha, \vw \right)/r_*  \\
\rb &=\rb\left(\alpha, \vw\right)\\
b &=b \left(\alpha, \vw\right),
\end{aligned} 
\een 
where $\hat{\Omega}_{\text{p}}$ is the maximum of $\OmGWSSM(z)$, 
$\zp$ is $z$ at the peak power and $\zb$ is the scale-free wavenumber at the second break. 
$J$ is the  timescale pre-factor defined in Eq.~(\ref{Eq:scaling_factors_GW}).

\begin{figure}[h!]
\centering 
\includegraphics[width=0.45\textwidth]{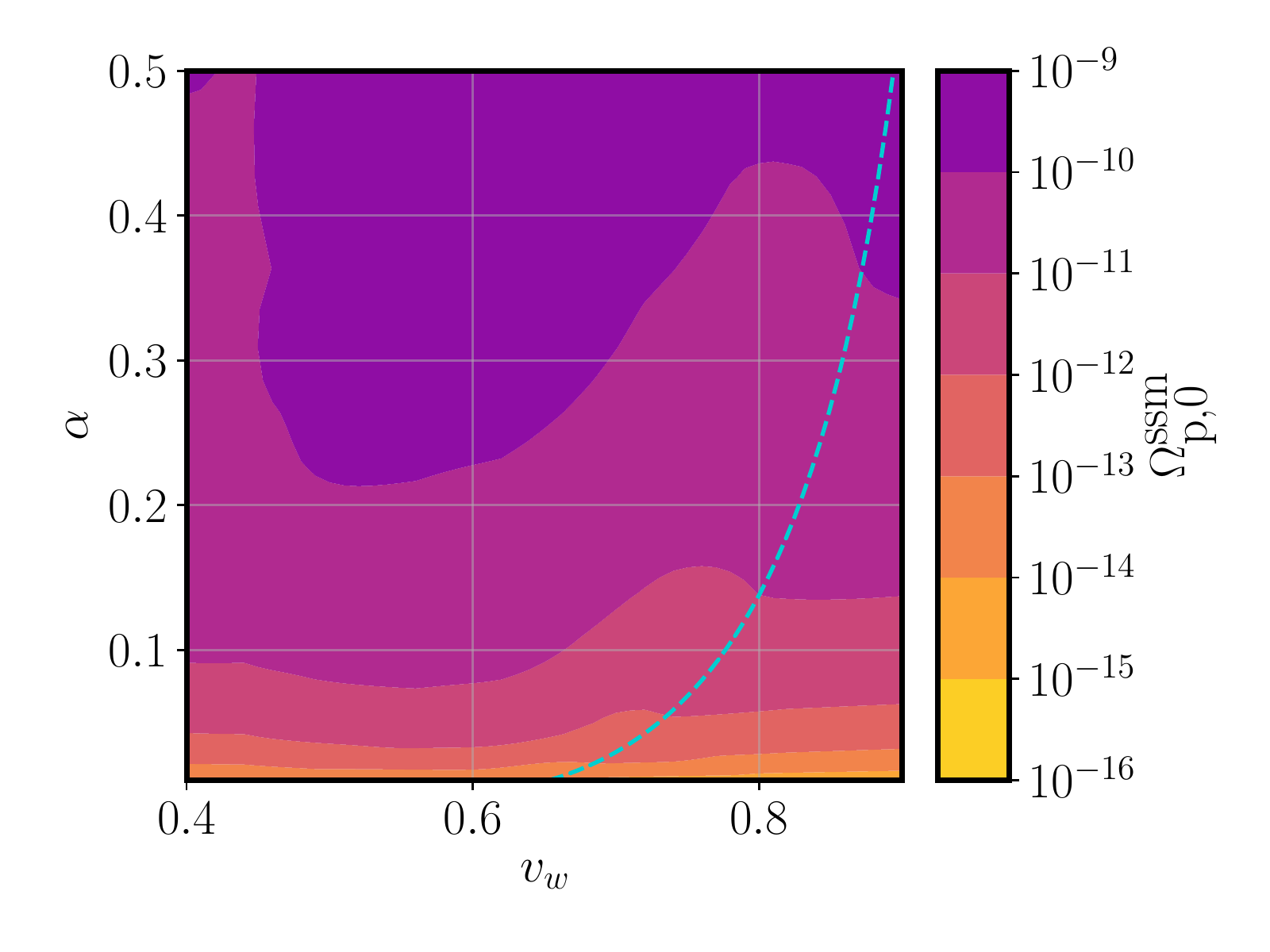}
\includegraphics[width=0.45\textwidth]{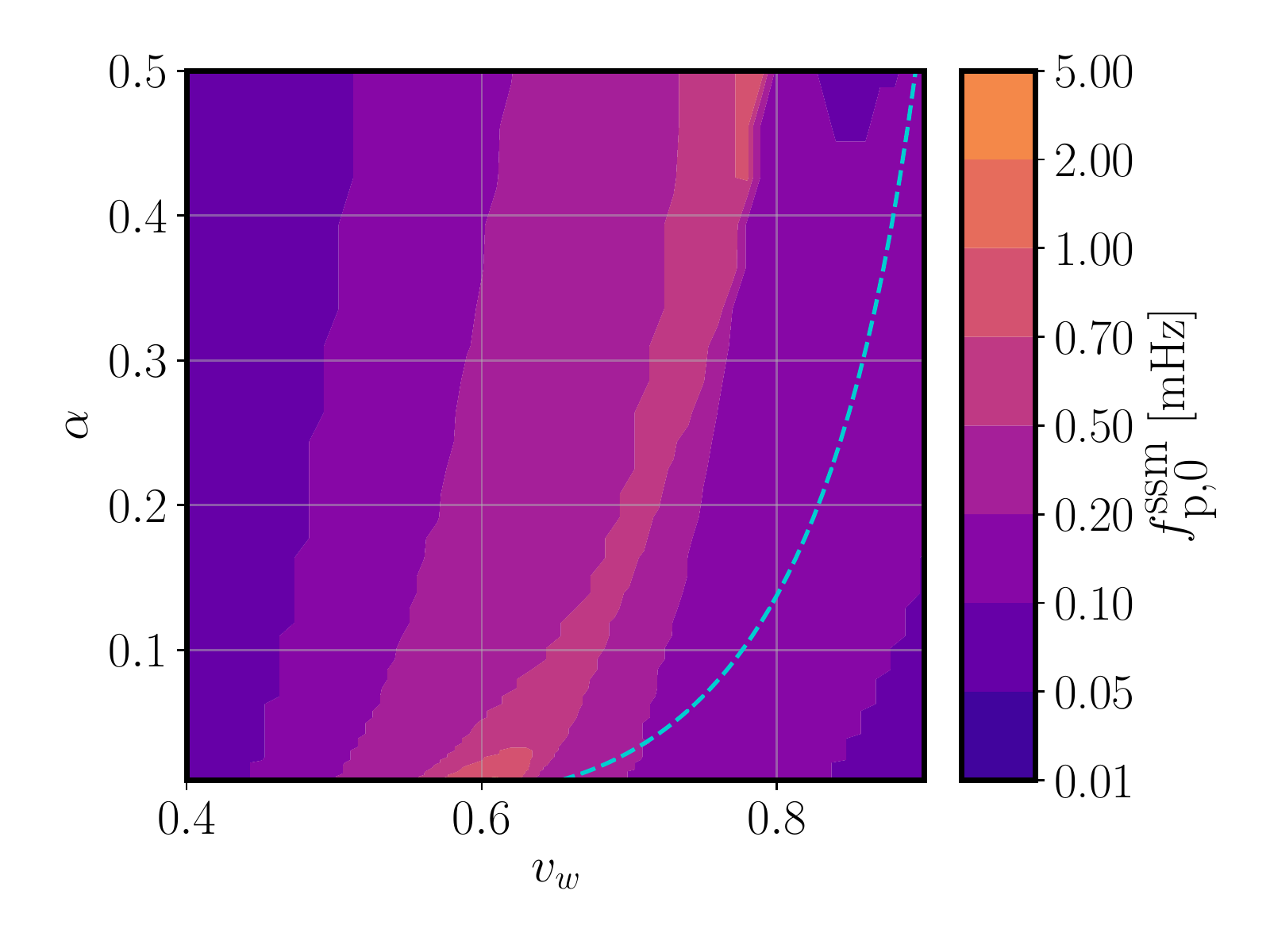}
\caption{ \label{fig:omp_hat_zp_hat_SSM} The peak power today $\Om_{\text{p,0}}^{\text{ssm}}$ and the peak frequency today $f_{\text{p,0}}^{\text{ssm}}$ caluclated with the sound shell model, for a range of wall speeds, $\vw$, and phase transition strengths,$\al$, The Hubble-scaled mean bubble spacing  $r_* = 0.1$ and nucleation temperature $\TN$. The turquoise dashed line shows the Jouguet speed Eq.~(\ref{Eq:Jouguet}).}
\end{figure}

\begin{figure}[h!]
\centering 
\includegraphics[width=0.45\textwidth]{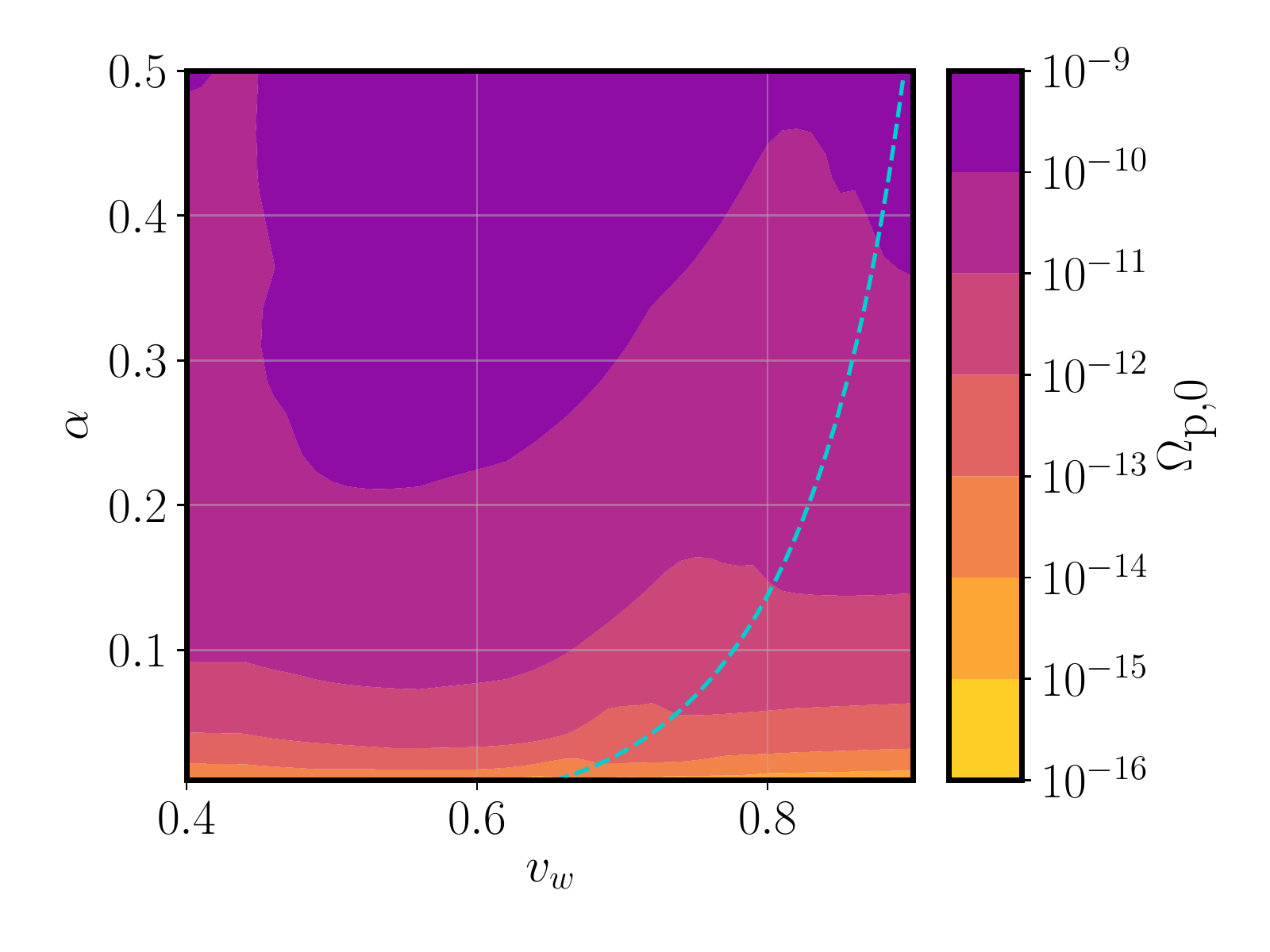}
\includegraphics[width=0.45\textwidth]{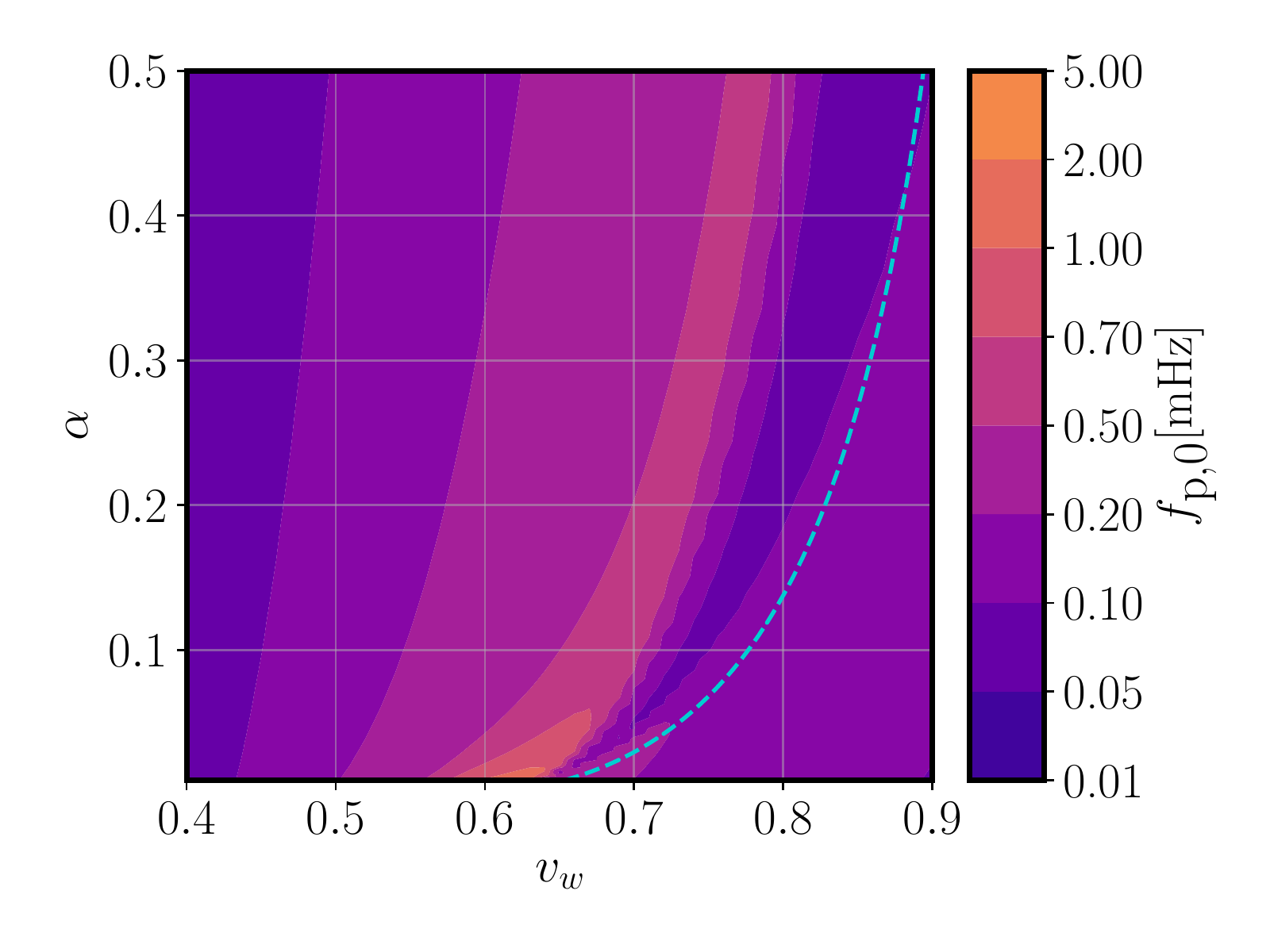}
\includegraphics[width=0.45\textwidth]{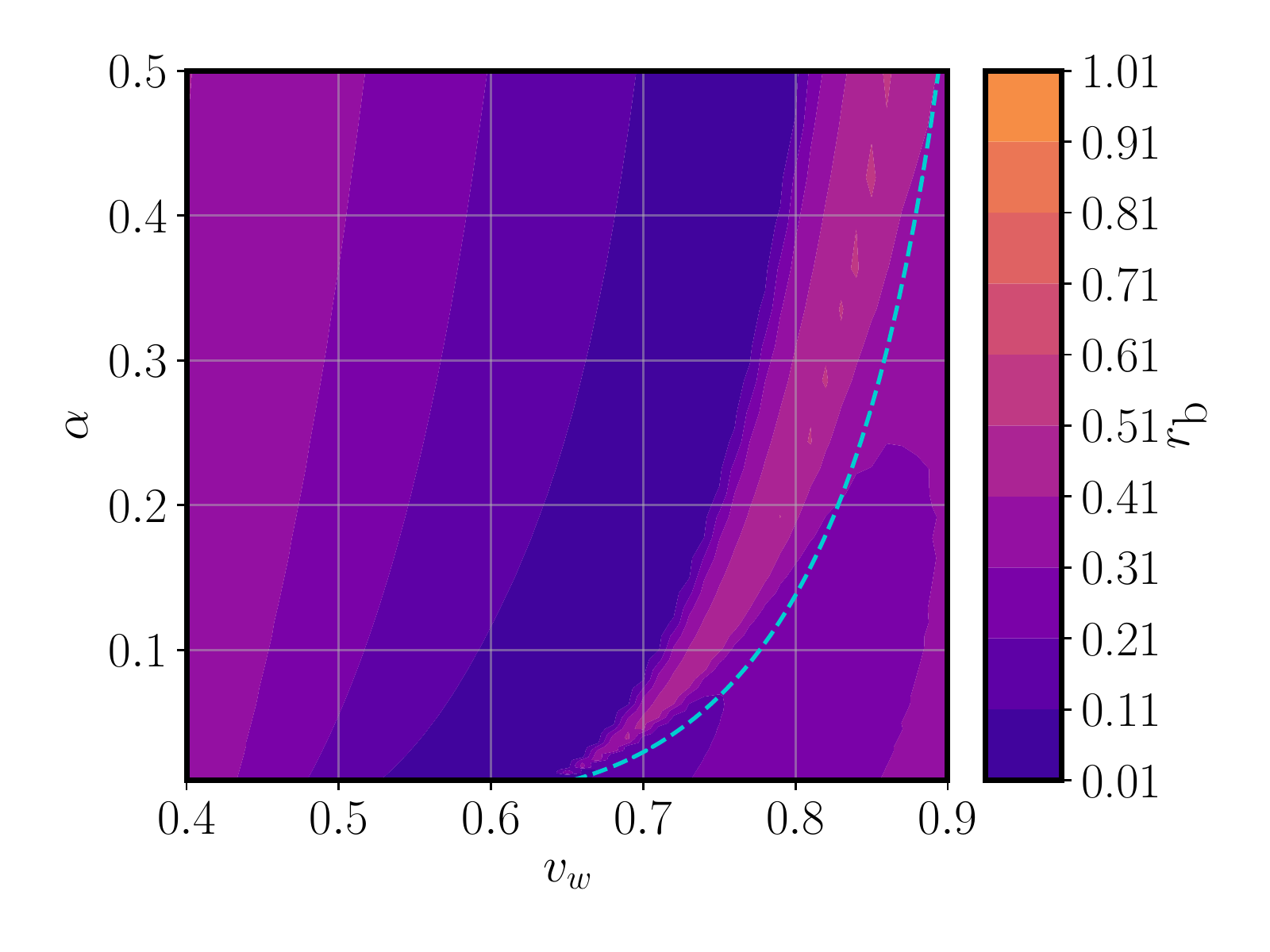}
\includegraphics[width=0.45\textwidth]{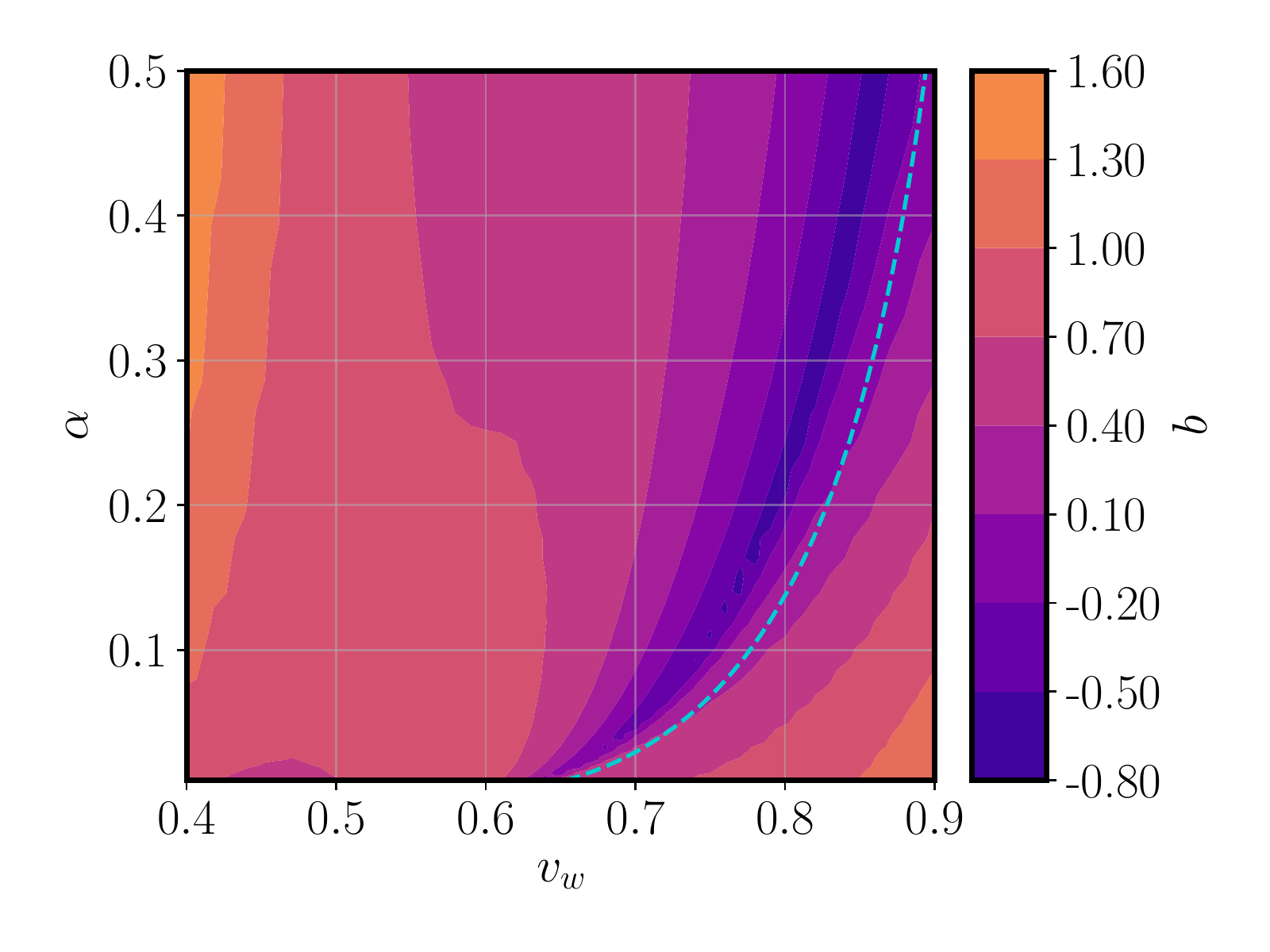}

\caption{\label{fig:4_param_contours} The best fit spectral parameters from fitting the double broken power law model to the power spectra from the sound shell model Eq.~(\ref{Eq:Omgw0_sup}).  $\Omega_{\text{p,0}}$ is the peak power today  with the Hubble-scaled mean bubble spacing $r_* = 0.1$ and $\Tn = 100$ GeV, $\fpt$ is the corresponding position of the peak (scaled using Eq.(\ref{Eq:spectral_parameters_thermo})), $\rb$ is the ratio of the frequency positions of the two breaks in the spectrum and $b$ the spectral slope of the power law between the two breaks. The turquoise dashed line is the Jouguet speed, Eq.~(\ref{Eq:Jouguet}).}
\end{figure}

 In Fig.~\ref{fig:omp_hat_zp_hat_SSM} we show the peak power today  $\Om_{\text{p,0}}^{\text{ssm}}$ and  the corresponding peak frequency $f_{\text{p,0}}^{\text{ssm}}$ calculated in  the SSM, using Eq.(\ref{Eq:Omgw0_sup}), for the $\vw$ and $\al$ parameter space of interest, with  $r_* =0.1$ and $\TN = 100$  GeV. Fig.~\ref{fig:4_param_contours}  shows the corresponding best fit spectral parameters for the double broken power law model. 


A comparison of the quality of the fit of the single broken power law and double broken power law models to the GW power spectrum from a first order phase transition, as described by the sound shell model, can be found in Appendix \ref{sec:fit_compare}.
\section{Noise model}
\label{sec:noise}
\subsection{LISA sensitivity curve}
The sensitivity of a gravitational wave detector can be characterised by the effective noise power spectral density\footnote{In this paper we consider two-sided power spectral densities, meaning the frequencies range from $- f_{\text{max}}$ to $+f_{\text{max}}$ } (PSD) $S(f)$, which is the gravitational strain 
spectral density required to produce a signal equal to the instrument noise $N(f)$.
If $\mathcal{R}(f)$ is the detector response function for gravitational waves,  
\begin{equation}
    S(f) \equiv \frac{N(f)}{\mathcal{R}(f)} .
\end{equation}{\label{Eq:S_def}}

LISA \cite{Audley:2017drz} is designed to be a triangular constellation of spacecraft connected by three pairs of laser links, through which changes in the distance between three pairs of free-falling test 
masses can be measured. The changes in distance are monitored through the differences in phase between the local oscillators and the remote spacecraft oscillators, communicated by the laser. The phase differences can be combined in different ways with different time delays 
to eliminate the laser noise \cite{Tinto:2001ii,Estabrook:2000ef}, by using the technique of time delay interferometry (TDI).

We work with the three noise-orthogonal TDI variables $\A$, $\E$ and $\T$ as described in \cite{Tinto:2002de,Prince:2002hp}. 
The $\T$ channel is insensitive to GWs at low frequencies. We will make the simplifying assumption that the $\T$ channel allows us to completely characterise the instrument noise.  

The instrument noise in LISA is expected to be dominated by two main sources: the test mass acceleration noise (acc), due to local disturbances of the test mass,
and the optical metrology noise (oms) which includes shot noise. 
As outlined in the LISA Science Requirements Document \cite{LISA_SR_doc} the target for the single link optical path-length fluctuations is
\ben{\label{Eq:P_oms}}
\Poms(f)=\left(\frac{1.5 \times 10^{-11} \mathrm{m}}{L}\right)^{2}\mathrm{Hz}^{-1},
\een
where $L = 2.5\times10^8\;\text{m}$ is the constellation arm length.  
The single test mass acceleration noise target is 
\ben{\label{Eq:P_acc}} 
\Pacc(f) = \left(\frac{3 \times 10^{-15} \,\text{m} \,\text{s}^{-2}}{(2 \pi f )^2L}\right)^{2}\left(1+\left(\frac{0.4 \mathrm{mHz}}{f}\right)^{2}\right) \mathrm{Hz}^{-1}.
\een
In the $\A$ and $\E$ channels the instrument noise is then (see e.g. \cite{Smith:2019wny})
\ben
\NA = \NE = \left[\left(4+2 \cos \left(f / \ft \right)\right) \Poms+8\left(1+\cos \left(f / \ft\right)+\cos ^{2}\left(f / \ft\right)\right) \Pacc\right]|W|^{2},
\een\label{Eq:Na}
where $\ft = c/(2 \pi L)$ is the transfer frequency and $c$ is the speed of light.
The function $W (f,\ft)= 1 - \exp(-2if/\ft) $ is the modulation caused by one round trip of a signal along a link. 
We use a simplified version of Eq.~(\ref{Eq:Na}) with $\cos\left(f / \ft\right) = 1 $, 
\ben \label{Eq:Na_simp}
\NA(f) \simeq \left(  6 \Poms(f) + 24\Pacc(f)\right)|W(f)|^{2} ,
\een 
which gives a reasonable fit to the true sensitivity curve. 

The gravitational wave response function for the $\A$ and $\E$ channels is known only numerically,\footnote{Recently, an analytic expression for the 
response function in the TDI $X$ channel has been derived \cite{Zhang:2020khm}, but 
the $\A$ and $\E$ channels also require the response function of the $XY$ cross-correlation. 
} 
but an approximate fit is 
\ben\label{Eq:Ra_fit} 
\mathcal{R}_{\A}^{\mathrm{Fit}} =\mathcal{R}_ {\E}^{\mathrm{Fit}}\simeq \frac{9}{20}|W|^{2}\left[1+\left(\frac{f}{4 \ft/3}\right)^{2}\right]^{-1}.
\een

We can now construct the  approximate noise power spectral density for the $\A$ and $\E$ channels using Eqs.~(\ref{Eq:S_def}), (\ref{Eq:Na_simp}) and (\ref{Eq:Ra_fit}):
\ben\label{Eq:SA}
S_{\text{A}}=  S_{\text{E}} =\frac{\NA}{\mathcal{R}_{\A}} \simeq \frac{40}{3}\left(\Poms + 4\Pacc \right)\left[1+\left(\frac{f}{4 \ft/ 3}\right)^{2}\right],
\een  
in this work we will be interested in the sensitivity to the GW fractional energy density power spectrum, which is related to the PSD by  
\ben{\label{Eq:OmInst}}
\OmInst =\left(\frac{4 \pi^{2}}{3 H_{0}^{2}}\right) f^{3} \SA(f)
\een 
we will refer to this as the LISA instrument noise  

\subsection{Extragalactic compact binaries}

A stochastic gravitational wave background (SGWB) from a superposition of unresolved extragalactic compact binaries is expected in the millihertz GW frequency band \cite{Regimbau:2011rp}. This signature is expected to be stationary, Gaussian and isotropic, distinguishable only by its frequency spectrum from cosmological signatures, such as a SGWB from a first order phase transition. 
It is composed of signals from stellar origin black hole binaries, neutron star binaries, 
and white dwarf binaries. 
These objects include precursors to compact binary mergers seen by the LIGO-Virgo collaboration \cite{LIGOScientific:2019vic}. 
We will refer to this as the extragalactic binary foreground (eb), 
which has the GW power spectrum 
\ben{\label{Eq:Om_LV}}
\Omeb(f)  = \Om_\text{ref,eb}\left(\frac{f}{f_\text{ref,eb}}\right)^{\frac{2}{3}} .
\een
We will assume that it is dominated by stellar origin black hole binaries, and take
$\Om_{\textrm{ref,eb}}$ to be the energy density of the LIGO-Virgo compact binaries at the reference frequency $f_{\textrm{ref,eb}} = 25 $ Hz.  
The current estimate is $\Om_{\textrm{ref,eb}} = 8.9 ^{+12.6}_{-5.6} \times 10^{-10}$  \cite{LIGOScientific:2019vic}. 
It is well below the instrument noise, and therefore not a significant contributor to the overall noise
relevant for stochastic backgrounds. This foreground is shown in Fig.(\ref{Fig:parameter_vary}). 
The contribution to the 
amplitude $\Om_{\textrm{ref,eb}}$ from black hole and neutron star binaries
will be more accurately measured by LIGO/Virgo once it reaches design 
sensitivity \cite{Martynov:2016fzi,TheLIGOScientific:2016wyq}, 
and by future ground-based detectors that may be online at a similar time to LISA \cite{Maggiore:2019uih,Sathyaprakash:2012jk}. 

\subsection{Unresolved galactic compact binaries }
A significant noise source for LISA is due to the large number of white dwarf binaries 
located within our galaxy \cite{Bender:1997hs,Evans:1987qa}. 
Some loud binaries will be individually resolvable, and as the mission progresses more will be identified.  At any mission stage, 
unresolved binaries will produce a confusion noise, which can be estimated using an iterative subtraction procedure outlined in \cite{Cornish:2017vip}. 
After a 4-year mission, estimates suggest around $20,000$ of the estimated $20$ million galactic binaries (gb) will be resolved, 
leaving a foreground\footnote{In this work we use the correction to the sign of the coefficient $b$ given in \cite{Schmitz:2020rag}.} 
\ben{\label{Eq:S_gb}}
    S _ {\text{gb}} ( f ) = A \left(\frac{1\,\textrm{mHz}}{f}\right) ^ { - 7 / 3 } \exp \left(  - \left(\frac{f}{f_{\text{ref,gb}}}\right) ^ { a } - b f \sin ( c f ) \right) \left[1 + \tanh \left( d \left( f _ { k } - f \right) \right)\right],
\een
where $A = 9 \times 10^{-38}$ $\textrm{mHz}^{-1}$ and $f_{\text{ref,gb}} =1000$ mHz. 
 The parameters $a$, $b$, $c$ and $f_k$ depend on the observation time: for a 4-year observation period, $a =0.138 $, $b =-0.221\, \text{mHz}^{-1} $, $c = 0.521 \,\text{mHz}^{-1} $, $d =1.680\; \text{mHz}^{-1} $ and the frequency of the knee of the power spectrum is $f_k = 1.13$ mHz. $S _ {\text{gb}} $ can be expressed in terms of energy density,
 \ben{\label{Eq:Om_gb}}
\Omgb = \left(\frac{4 \pi^{2}}{3 H_{0}^{2}}\right) f^{3} S_{\textrm{gb}}(f) ,
\een
which we will refer to as the galactic binary foreground,  and show in Fig.~\ref{Fig:parameter_vary}. There is potential for this foreground to be extracted separately,  due to the annual modulation in the signal as LISA's direction of maximum sensitivity sweeps past the galactic plane \cite{Adams:2013qma}.
If no attempt to remove the annually modulated stochastic signals is made, galactic binaries will be a significant source of noise around 1 mHz.
We will consider parameter sensitivity both with and without the galactic binary foreground, to estimate upper and lower bounds. 

In addition to the foregrounds considered here there are a number of other sources that may need to be considered when trying to extract a stochastic GW background from a first order phase transition. These include confusion noise from unresolved extreme mass ratio inspirals  \cite{Bonetti:2020jku} and a foreground from unresolved massive black hole binaries \cite{Sesana:2004gf}. 
In addition, extragalactic white dwarf binaries could contribute significantly to the compact binary foreground \cite{Farmer:2003pa}. 
We choose to leave the inclusion of these foregrounds for future work:  current models are not as well characterised, 
and, at least in the case of massive black hole binaries, are expected to be less significant. 

\subsection{Signal-to-noise ratio}
\begin{figure}[ht!]
\centering 
\includegraphics[width=0.8\textwidth]{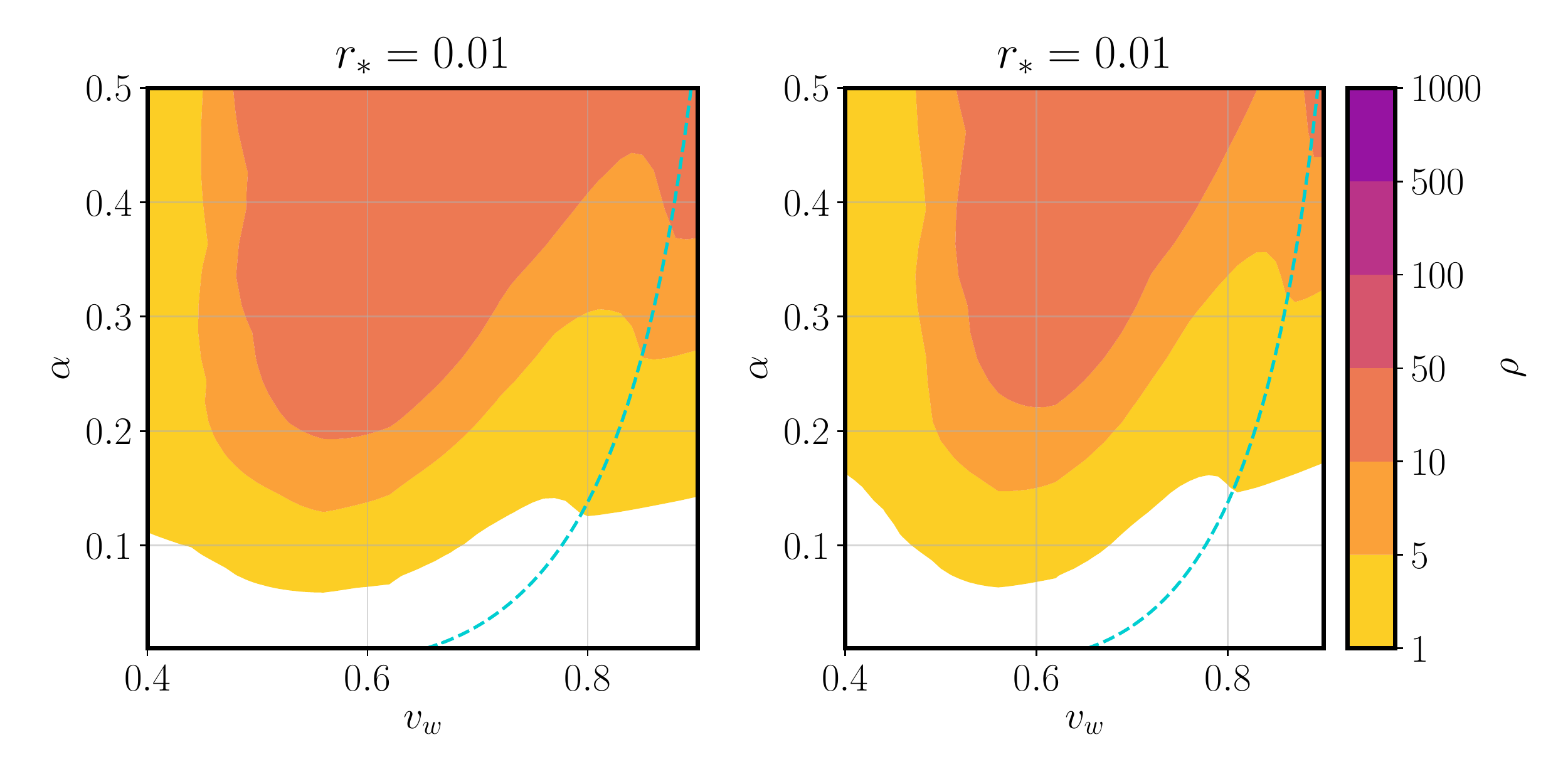}
\includegraphics[width=0.8\textwidth]{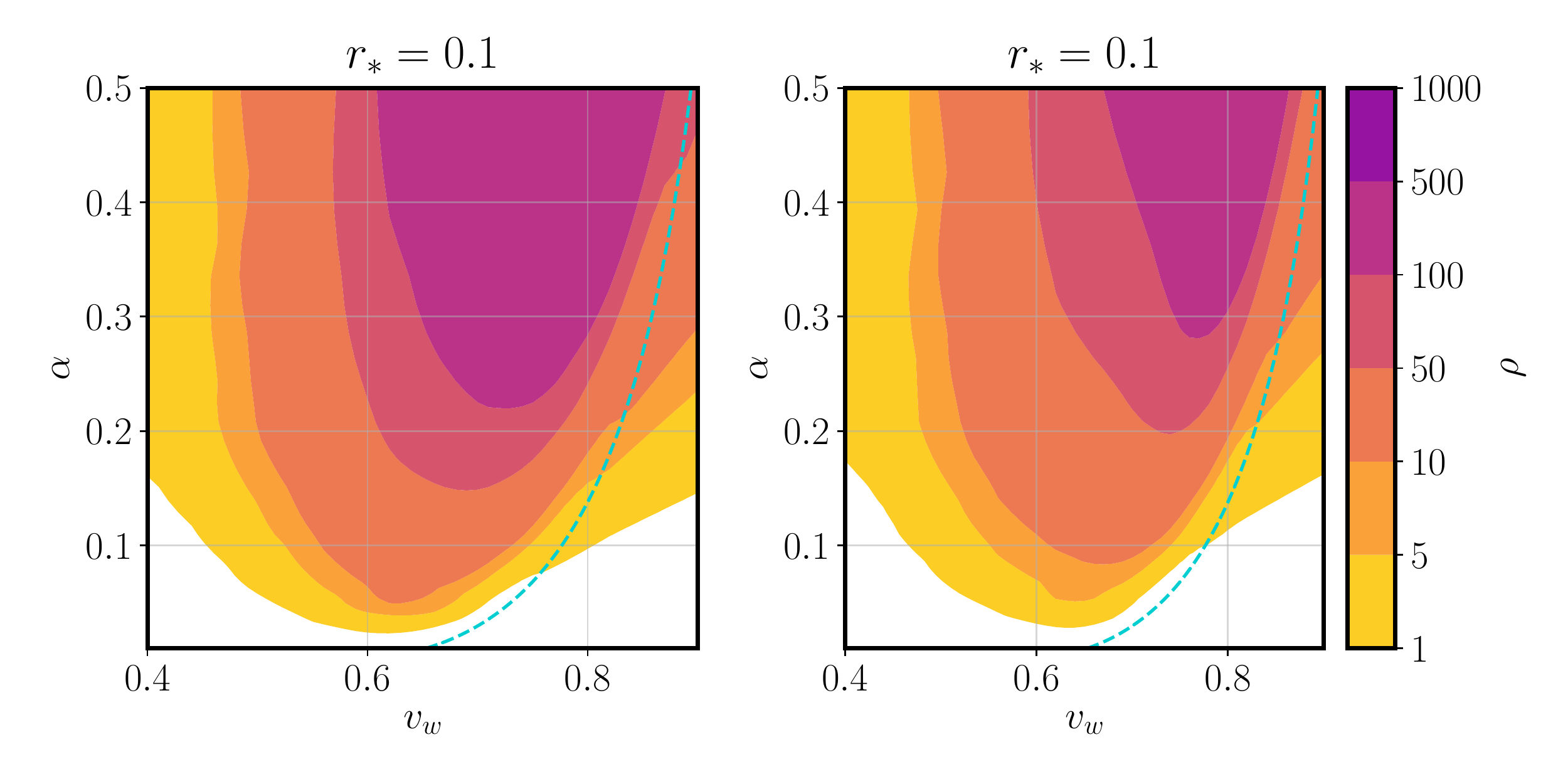}
\caption{\label{fig:SNRs} The signal-to-noise $\SNR$ for different combinations of the wall speed $\vw$, phase transition strength $\al$, Hubble-scaled mean bubble spacing $r_*$, with the nucleation temperature $\TN = 100$ GeV. 
In the left column the noise model includes the LISA instrument noise - Eq.~(\ref{Eq:OmInst}), the foreground from unresolved stellar origin black hole binaries - Eq.~(\ref{Eq:Om_LV}). In the right hand column we also include the unresolved galactic binary foreground- Eq.~(\ref{Eq:Om_gb}). The turquoise dashed line shows the Jouguet detonation speed, the minimum speed of a detonation for each $\al$, given in Eq.~(\ref{Eq:Jouguet}).  }
\end{figure}

As a first assessment of whether a signal is observable or not, one can calculate the signal-to-noise ratio $\SNR$ by comparing the signal $\OmGW$ with the noise model $\Om_\text{n}$ \cite{Allen:1997ad,Maggiore:1999vm}:
\ben{\label{Eq:SNR}}
\SNR = \sqrt{\Tobs\int^{f_{\text{max}}}_{f_{\text{min}}} df \frac{h^2\OmGWt^2}{h^2\Om_\text{n}^2}},
\een
where $\Om_\text{n}$ is the sum of all sources of noise. Our base noise model consists of the LISA instrument noise as given in  Eq.~(\ref{Eq:OmInst}), the extragalactic background Eq.~(\ref{Eq:Om_LV}) and the galactic binaries Eq.~(\ref{Eq:Om_gb}), so that 
\ben\label{Eq:base_noise}
\Om_\text{n} = \OmInst + \Omeb + \Omgb.
\een
In this work we take the observation time $\Tobs = 4$ years, which is LISA's designed mission lifetime \cite{Audley:2017drz}, so that we can use our noise model in combination with the prediction of the galactic binary foreground given in \cite{Cornish:2017vip}. 
LISA's science operational time is expected to be $\approx 75\%$ to the total mission lifetime, but the mission may last up to 10 years. 

In Fig.~\ref{fig:SNRs}, we calculate $\SNR$ for the thermodynamic parameter space explored in this paper. The power spectra are described by Eq.~(\ref{Eq:Omgw0_sup}). Generally, $\SNR$ is larger for stronger phase transitions, corresponding to larger $\al$, and sensitivity to the wall speeds $\vw$ peaks in the region of the speed of sound $\cs$. 

\section{Fisher matrix analysis}
\label{sec:fm}

An estimation of LISA's sensitivity to parameters that describe a first order phase transition can be obtained by Fisher matrix (FM) analysis. The FM is 
the curvature of a Gaussian approximation to the posterior likelihood around the maximum. 
The inverse of the FM is the covariance matrix, the diagonal elements of which give an approximation of the uncertainty in any given parameter.  

\subsection{LISA likelihood model}

Here we outline how we model the LISA data, explaine the assumptions made, and define the likelihood used. 
We model the LISA data as an unbroken  $\Tobs =  4$ yr stream with a regular data sampling interval $ T_{\text{samp}}=5$ s. 
The frequency domain gravitational wave strain amplitude $h(f)$ is the Fourier transform of the strain time series $h(t)$:
\ben{\label{Eq:hf}} 
h(f_n) = 
\frac{1}{\sqrt{N}}\sum_{m = 0}^{N-1} h(t) \exp{\left(- 2 \pi i f_n t_m \right )},
\een
where $t_m = m T_{\text{samp}}$ and $f_n = n /\Tobs $, 
with $-N/2 < n \le N/2$, and $N =  \Tobs/ T_{\text{samp}}$. 
The strain amplitude is related to the gravitational wave power spectrum by 
\ben
\OmGWt(f_n) =  \left(\frac{4 \pi^{2}}{3 H_{0}^{2}}\right) f_n^{3} {|h(f_n)|^2}.
\een
We consider the time series of $A$ and $E$ TDI channels, which in the frequency domain have  
power spectral densities  $\DAn =|\A(f_n)|^2$ and $\DEn = |\E(f_n)|^2$. The variances of the Fourier amplitudes in the $A$ and $E$ channels are taken to be identical and independent, and written $\Sn$, and to depend on a vector of model parameters $\vec{\th}$. 

As $N \simeq 2\times 10^7$ it saves computation time to group the data by frequency binning, which is a crude way 
of grouping the data, but sufficient for the level of analysis we carry out.
We split the frequency range into a set of $\Nbin$ logarithmically spaced positive frequency bins, with
bin boundaries $\fbin$, where $b $ ranges from $0$ to $\Nbin$. In each bin there are 
\ben\label{Eq:nb}
\nbin =  \left[ (f_{b } - f_{b-1}) \Tobs \right]
\een
different frequencies, where the square brackets denote the integer part. 


First considering $\DAn$ we define $\DAbin$  to be the weighted mean value for $D^{\A}_n$ in bin $b$, so that 
\ben
\DAbin =  \frac{S_b}{n_b} \sum_{n \in I_b} \frac{\DAn}{S_{n}} ,
\een
with  
\ben 
\frac{1}{S_{b}}  = \frac{1}{n_b} \sum_{n \in I_b} \frac{1}{S_{n}} ,
\een 
where $I_b$ the set of integers $n$ such that $|f_n|$ is in frequency bin $b$ and $S_n$ is the mean value of $D^{\A}_n$.

As $\DAbin$ is the average of squares of $2n_b$ normally distributed real random variables, 
the likelihood is a chi-squared distribution \footnote{In this paper we echo the notation used in \cite{Pieroni:2020rob}.}
\ben
p\left(\DAbin \mid S_{b}\right)=\prod_{b=1}^{N_{b}} \frac{1}{\left(n_{b}-1\right) !} \frac{n_{b}}{S_{b}}\left(n_{b} \frac{\DAbin}{S_{b}}\right)^{n_{b}-1} \exp \left(-n_{b} \frac{\DAbin}{S_{b}}\right).
\een
It will be convenient to approximate the likelihood with a Gaussian distribution, 
using the central limit theorem with the assumption $n_b \gg 1$  for all bins. The distribution  
for $\DAbin$ has mean $\Sbin$ and 
variance $ \Sbin^2/\nbin $, giving the Gaussian approximation 

\ben\label{Eq:Likeli_CLT} 
p(\DAbin| \Sbin) = \prod_{b=1}^{\Nbin}  \left(\frac{\nbin}{2\pi \Sbin^2}\right)^{\frac{1}{2}} \exp{\left(-\frac{1}{2}\frac{\nbin\left(\DAbin  - \Sbin\right)^{2}}{\Sbin^2}\right)}.
\een 
As the $\A$ and $\E$ channels are uncorrelated and are assumed to a first approximation have identical noise, 
we can combine them into an average variable $\Dbin = \left(\DAbin +\DEbin \right)/2$ with variance $\Sbin^2/2\nbin$.
The likelihood for the binned average spectral density $\Dbin$ is then 
\ben\label{Eq:Likeli_CLT_A_E} 
p(\Dbin| \Sbin) = \prod_{b=1}^{\Nbin}  \left(\frac{2\nbin}{2\pi \Sbin^2}\right)^{\frac{1}{2}} \exp{\left(-\frac{1}{2}\frac{2\nbin\left(\Dbin  - \Sbin\right)^{2}}{\Sbin^2}\right)}.
\een 
The Gaussian approximation is known to be biased  \cite{Bond:1998qg,Verde:2003ey,Hamimeche:2008ai}, 
and one could improve the accuracy of the likelihood with a log 
normal correction \cite{Verde:2003ey,Hamimeche:2008ai,Flauger:2020qyi}.
One can also evaluate the Fisher matrix directly with the chi-squared distribution. 
On the other hand, 
using the Gaussian approximation simplifies the calculations. As we now show,
we always have $2n_b \gtrsim \text{O}(10^2)$, for which the Gaussian approximation is sufficiently accurate.

Firstly, when working with the double broken power law model, we take $\Nbin = 100$ logarithmically spaced 
frequency bins, for which $\nbin \gtrsim 155$. The frequency binning for calculations in the sound shell model is a little more complicated, as 
we calculate the theoretical model in terms of the angular frequency scaled by the 
mean bubble separation $z = k\Rstar$, rather than an absolute frequency. 
This avoids having to recompute power spectra for different $r_*$ and $\TN$,  
as the shape of the GW power spectrum depends only on the thermodynamic parameters $\vw$ and $\al$.

The transformation from $z$ to a frequency today is 
\ben
f = f_{*,0} z /r_* ,
\een
where $f_{*,0}$ is given in Eq.~(\ref{Eq:f0}),
and we recall that $r_* = \Rstar \HN$. 
In this case, the number of data frequencies in bin $b$ is 
 \ben
 \nbin =  f_{*,0} \Tobs \Delta z_b /r_*
 \een
where $\Delta z_b = z_{b} - z_{b-1}$. 
In this work we compute  $100$ $z $ values with logarithmic spacing between $1$ and $1000$. 
For the LISA data described here,  the minimum $\nbin \simeq 42 $.

The Fisher matrix is 
\ben
F_{ij} = \left\langle \frac{\partial l_G}{\partial\theta_i}\frac{\partial l_G}{\partial\theta_j}\right\rangle,
\een
where $\theta_i$ denotes the $i$th component of the vector of model parameters $\vec{\th}$, 
and  the Gaussian approximation to the log-likelihood is 
\ben 
l_G = \ln(p) = -\frac{1}{2} \sum_{b=1}^{\Nbin} \frac{2\nbin\left(\Dbin  - \Sbin\right)^{2}}{\Sbin^2} - \sum_{b=1}^{\Nbin} \ln \Sbin+ \text{const} .
\een 
Hence the Gaussian approximation to the Fisher matrix is 
\ben
F_{ij}^{G} =  \sum_{b=1}^{\Nbin} \frac{2\nbin}{\Sbin^2} \frac{\partial \Sbin}{\partial\theta_i}\frac{\partial \Sbin}{\partial\theta_j} .
\een

One can show that the Fisher matrix calculated using the full chi-squared distribution is 

\ben
F_{ij} = \left(1+\frac{1}{2 \nbin}\right)F_{ij}^{G},
\een
Hence with smallest value of $2\nbin = 84$, the difference is minimal, and 
we take $F_{ij}$ to be its Gaussian approximation from now on.  

We will also use the power spectra, $\Omega_{\textrm{t}}(f_b,\vec{\theta}) $, rather than the spectral densities $\Sbin$ to formulate the theoretical model of the data:
\ben{\label{Eq:Omn}}
\Omega_{\textrm{t}}(f_b,\vec{\theta}) = \Omega_{\textrm{n}}(f_b) + \Om_{\textrm{fg}}(f_b) + \Om_{\textrm{pt}}(f_b,\vec{\th}) ,
\een
where we assume that the instrumental noise $\Omega_{\textrm{n}}$ and foregrounds $\Om_{\textrm{fg}}$ are much 
better known than parameters of the phase transition signal $\Om_{\textrm{pt}}$, 
and so the parameters in the Fisher matrix are just those describing the phase transition. 
We consider two kinds of foregrounds: one from extragalactic binaries Eq.~(\ref{Eq:Om_LV}), and one with both extragalactic and galactic binaries Eq.~(\ref{Eq:Om_gb}).  As only the ratio of spectra appears, the Fisher matrix in terms of the power spectrum is simply 
\ben{\label{Eq:Fij}}
F_{ij} = \Tobs \sum_{b=1}^{\Nbin} \frac{2\Delta f_b }{\Omt^2} \frac{\partial \Omt}{\partial\theta_i}\frac{\partial \Omt}{\partial\theta_j} .
\een 
The sum on the right-hand side can be viewed as a numerical approximation to an integral over frequencies.

The covariance matrix is the inverse of the Fisher matrix, 
\ben\label{Eq:rel_u}
C_{ij} = F^{-1}_{ij} ,
\een
where the square roots of diagonal entries give the uncertainty in the $i$th parameter $\De \th_i$. 
These uncertainties include correlations between parameters. 
 We define $\th_i$ to be the logarithmic model parameter, so the square roots of the diagonal entries in the covariance matrix are the relative uncertainty in the parameter.

\subsection{Principal components}
The Fisher matrix can be used to construct a set of uncorrelated and orthonormal observables, the principal components. 
As the Fisher matrix is a symmetric $n \times n$ matrix we can find its eigenvectors and eigenvalues 
\ben\label{Eq:Fisher_eigen}
F = U\Lambda U^{\dagger}, \quad \Lambda = \text{diag}(\la_1,\la_2,\la_3,\la_4),
\een
where $U$ is a matrix of the orthonormal eigenvectors of the Fisher matrix, $u_n$ is the $n^{\text{th}}$  eigenvector and $\la_n$ is the $n^{\text{th}}$ eigenvalue. We can then construct a new set of variables $\vec{X} = (X_1,X_2,X_3,X_4)$ , each $X_n$ being a linear combination of our original parameters calculated by using $U^{\dagger}$ as a projection vector, 
\ben\label{Eq:principal_components}
\vec{X} =  U^{\dagger} \bar{\th}.
\een
$X_n$ is the $n^{\text{th}}$ principal component and the standard deviation in $X_n$ is $\la_n^{-1/2}$. The principal components are ordered according to the size of their corresponding eigenvalues, meaning $X_1$ is the highest-order and best constrained parameter, and $X_4$ is the lowest order parameter and worst constrained \cite{Efstathiou:1998xx}.

\section{Fisher matrix calculation and relative uncertainties}
\label{sec:results}
In this section we calculate the Fisher matrix and the relative uncertainties as described in the previous section, for several scenarios. Firstly, the FM for the spectral parameters from the double broken power law fit to the SSM. Then we evaluate the FM for the four key thermodynamic parameters in the SSM 
$(\vw,\al,r_*,\TN)$ for two cases: free and fixed nucleation temperature $\TN$. We also calculate the expected sensitivity to the principal components of the GW power spectrum calculated with the SSM. In each case we investigate the impact of including the foreground from galactic binaries. 

\subsection{Double broken power law model}

\begin{figure}[h!]
\centering 
 \includegraphics[width=1\textwidth]{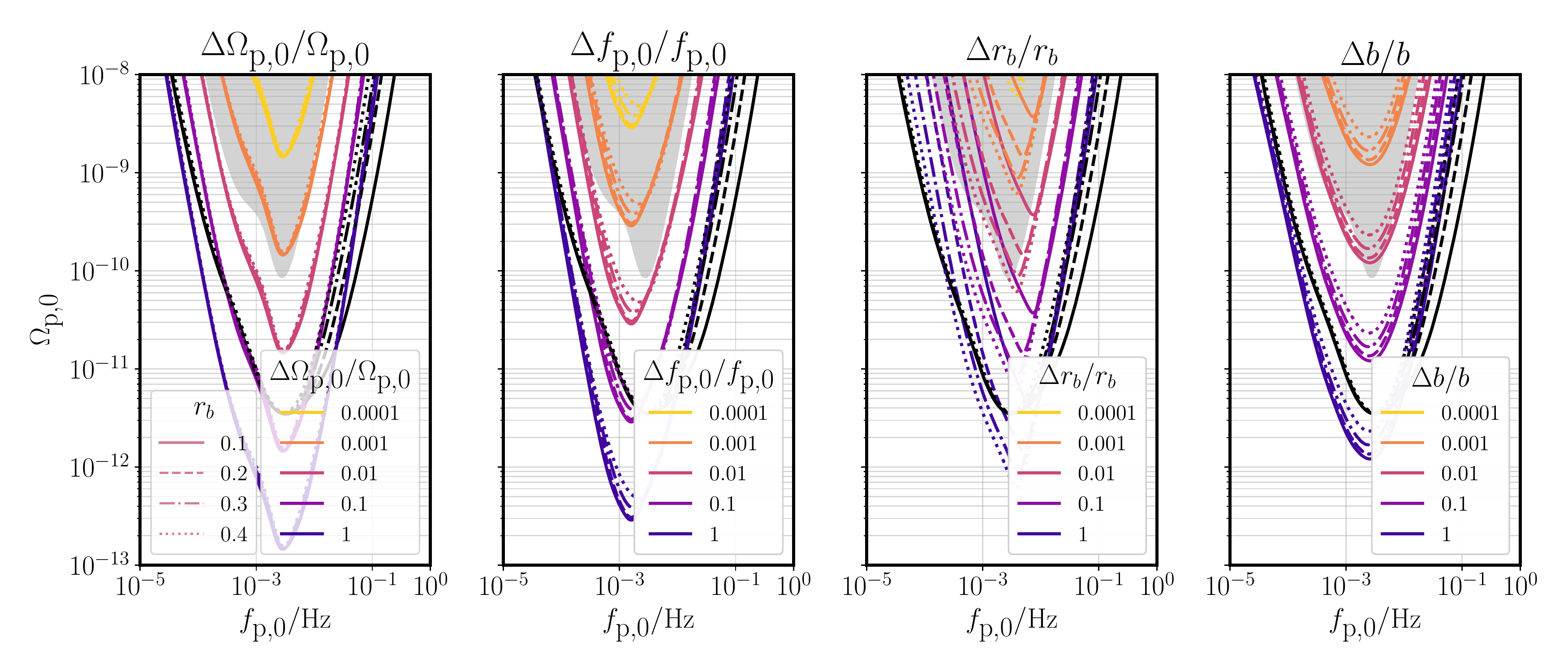}
 \includegraphics[width=1\textwidth]{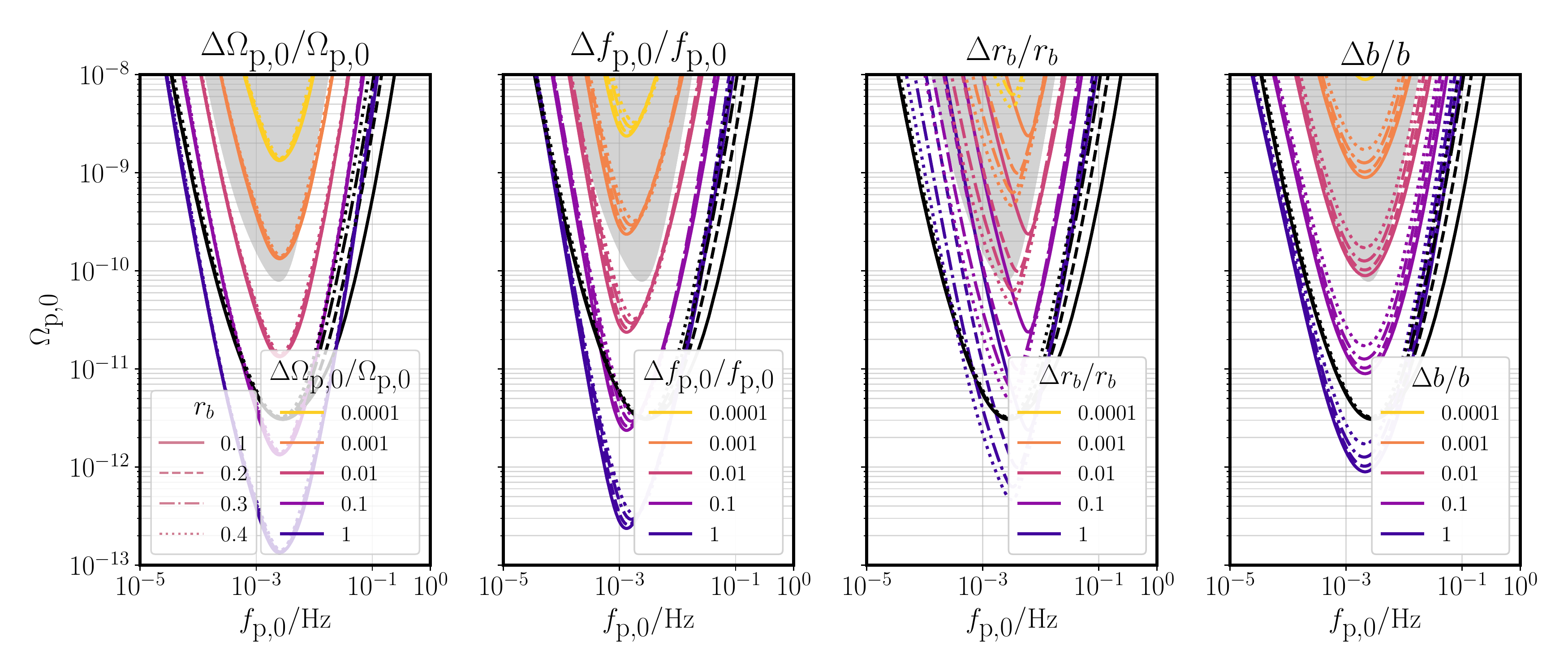}

\caption{\label{fig:obs_P_rel_uncer} 
Coloured contours show relative uncertainties calculated from the Fisher matrix for the 
parameters of the double broken power law model (Eq.~\ref{Eq:omgw_dbl_brkn}):
peak power $\OmPeakt$, peak frequency $\fpt$, break ratio $\rb$ and intermediate power law $b$. The line styles indicate the break ratio values $r_b$.  The black lines  show contours of signal-to-noise ratio $\SNR = 20 $ for different $\rb$, with the same line styles. 
The grey shaded area indicates the region where the peak signal power is above the combined instrumental noise and foregrounds. 
In the upper panel the noise model consists of the LISA instrument noise, Eq.~(\ref{Eq:OmInst}), foreground from compact binaries, Eq.~(\ref{Eq:Om_LV}) and the galactic binary foreground, Eq.~(\ref{Eq:Om_gb}). In the lower panel the galactic binary foreground is removed. }

\end{figure}

First we look at the relative uncertainty for the spectral parameters as described in the proposed double broken power law model given in Eq.~(\ref{Eq:omgw_dbl_brkn}). 
 In this case the parameters are 
 $\vec{\theta} = \left(\ln(\OmPeakt),\ln (\fpt),\ln(\rb), \ln(b)\right)$. The 
 Fisher matrix entries can be calculated analytically. The gravitational wave power spectrum is evaluated at $100$ frequencies with logarithmic spacing between $10^{-5}\;\text{Hz}$ and $1\;\text{Hz}$. 
We sample the parameter space as follows: 200 peak powers $\OmPeakt$, with logarithmic spacing between $ 10^{-13}$ and $ 10^{-8}$; 200 peak frequencies $ \fpt $, with logarithmic spacing between  $10^{-5}$ and $1$; 4 frequency break ratios $\rb = [0.1,0.2,0.3,0.4]$; and intermediate power law with spectral slope $b=1$, the generic value, as explained in \cite{Hindmarsh:2019phv}.

The range of spectral parameters at which we evaluate the relative uncertainties was chosen such that they could be produced by thermodynamic parameters currently explored in models and simulations (as displayed in Fig.~\ref{fig:4_param_contours}).  
The resulting relative uncertainties, with and without the foreground from 
unresolved galactic binaries, are shown  in Fig.~\ref{fig:obs_P_rel_uncer} as contours in the 
$(\OmPeakt,\fpt)$ plane. The line style shows the frequency break ratio $\rb$, and the colour 
the relative uncertainty.  Also plotted is the curve at which the signal-to-noise ratio $\rho$ is 20, 
and for comparison, the noise model (which includes foregrounds)  
$\Omega_{\textrm{n}}(f) $ as a black line.

It can be seen that $\rh=20$ is reached for peak powers well below the noise 
level, which is an effect of the integration over frequencies.  One can regard the $\rho=20$ line  
as a peak-integrated sensitivity \cite{Schmitz:2020rag}, 
which generalises the idea of power law sensitivity \cite{Thrane:2013oya} to 
peaked power spectra. 

 The results for the relative uncertainty in $\OmPeakt$ and $\fpt$ are consistent with 
 those in Ref.~\cite{Hashino:2018wee}, which studied the two-parameter single broken power law model 
 advocated by the LISA Cosmology Working group \cite{Caprini:2019egz}. 
 One can summarise the conclusion 
in a parameter-independent way by the statement that a SNR of about 20 allows a measurement of the 
 peak power and peak frequency at around a 10\% level of uncertainty. 
 If the unresolved galactic binaries are not 
 removed, the parameter space required to achieve $\rh=20$ is reduced.  
 
 A 10\% measurement of $\rb$, which encodes information about the wall speed, 
 requires higher signal-to-noise ratios, with the best resolved break ratio being $\rb = 0.4$.  
 This is the value of $\rb$ giving a power spectrum with the narrowest peak, 
and so the whole peak is likely to be in the sensitivity window of the detector. 

\subsection{Sound shell model}
In the simplest version of the sound shell model we study, the parameters are the logarithms of the 
wall speed, the phase transition strength, the Hubble-scaled mean bubble spacing and the nucleation temperature,
giving a parameter vector  
$\vec{\th} = (\ln \vw, \ln\al,\ln r_*, \ln(\TN/\text{GeV}))$.

We evaluate the Fisher matrix at all combinations of our parameter space using Eq.~(\ref{Eq:Fij}). 
The parameter space was sampled with 50 wall speeds $\vw$ in the range $0.4 \le \vw \le 0.9$, 51 phase transition strengths $ \al $ logarithmically spaced between   $0.01$ and $0.5$, 2 Hubble-scaled mean bubble spacings $r_* = 0.01,0.1$, and nucleation temperature $\TN = 100\;\text{GeV}$.

To construct the Fisher matrix we need to calculate the partial differentials of the GW power spectrum with respect to each of our thermodynamic parameters. 
The gradients with respect to $\vw$ and $\al$ were computed numerically.  
The derivatives with respect to the Hubble-scaled mean bubble spacing $r_*$ and the nucleation temperature $\TN$ are calculated as follows.

With the phase transition model spectrum $\Om_{\textrm{pt}}$ given by  $\OmGWt$ in Eq.~(\ref{Eq:Omgw0_sup}), 
we recall that 
\ben\
J = H_n R_* H_n\tau_v  = r_* \left(1 -  \frac{1}{\sqrt{1 + 2x}} \right), 
\een
where $x = r_*/\sqrt{K(\al,\vw)}$. The gravitational wave frequency today $f$ is related to the dimensionless wavenumber $z$ 
through $ z = r_*(f/f_{*,0})$, with the reference frequency depending on $\TN$ through Eq.~(\ref{Eq:f0}).
Hence we find 
\ben
\frac{\partial \Omega_{\textrm{t}}(f)}{\partial \ln r_*} =  \OmGWt \left( \frac{\pa \ln J}{\pa \ln r_*} + \ga_\text{gw}(z)\right),
\een
where 
\ben
 \frac{\pa \ln J}{\pa \ln r_*} =  1 + \frac{r_*}{J} \frac{x}{\left( 1 + 2x \right)^{3/2}},
\een
and 
$ \ga_\text{gw} = d \ln \OmGW/d\ln z$ is the local power law index of the gravitational wave power spectrum, 
which we compute numerically.

The partial differential with respect to  $\TN$ is then 
\ben
\frac{\partial \Omega_{\textrm{t}}(f)}{\partial \ln( \TN/\text{GeV})} =  - {\OmGWt}\ga_\text{gw}(z).
\een

\begin{figure}[h!]
\begin{subfigure}[b]{\textwidth}
\centering 
\rotatebox{90}{\hspace{20mm}$r_* = 0.01$}
\includegraphics[width=0.8\textwidth]{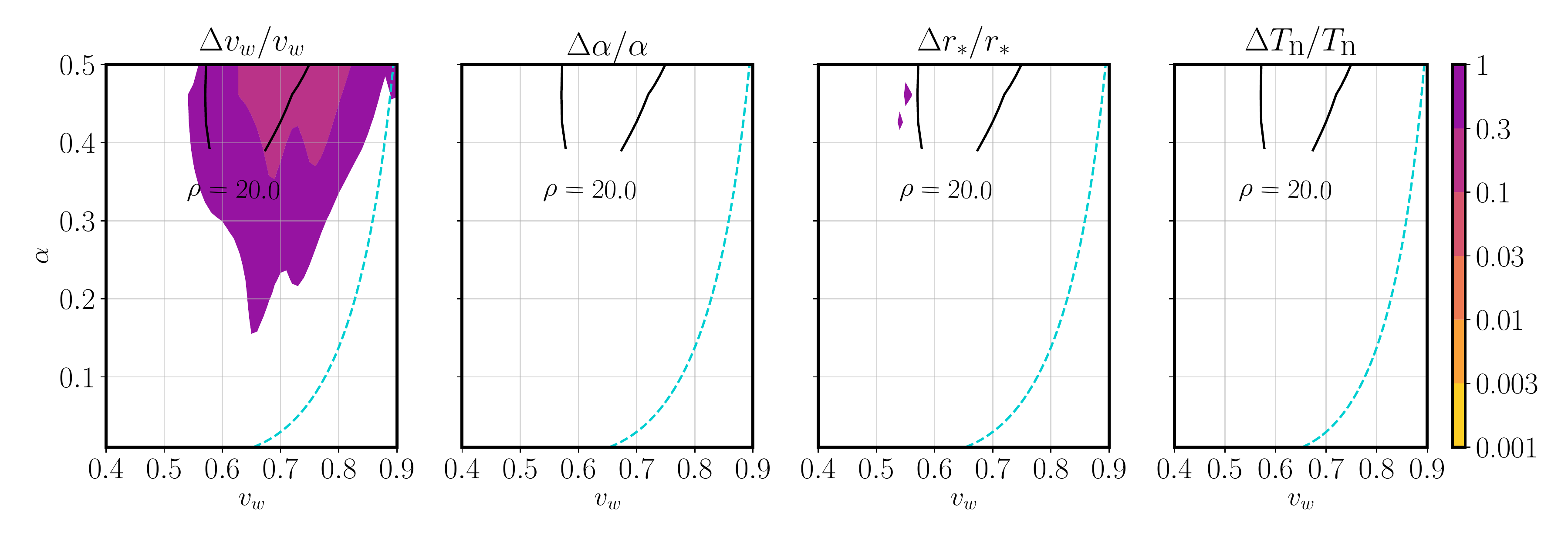}\\[-10pt]
\rotatebox{90}{\hspace{20mm}$r_* = 0.1$}
\includegraphics[width=0.8\textwidth]{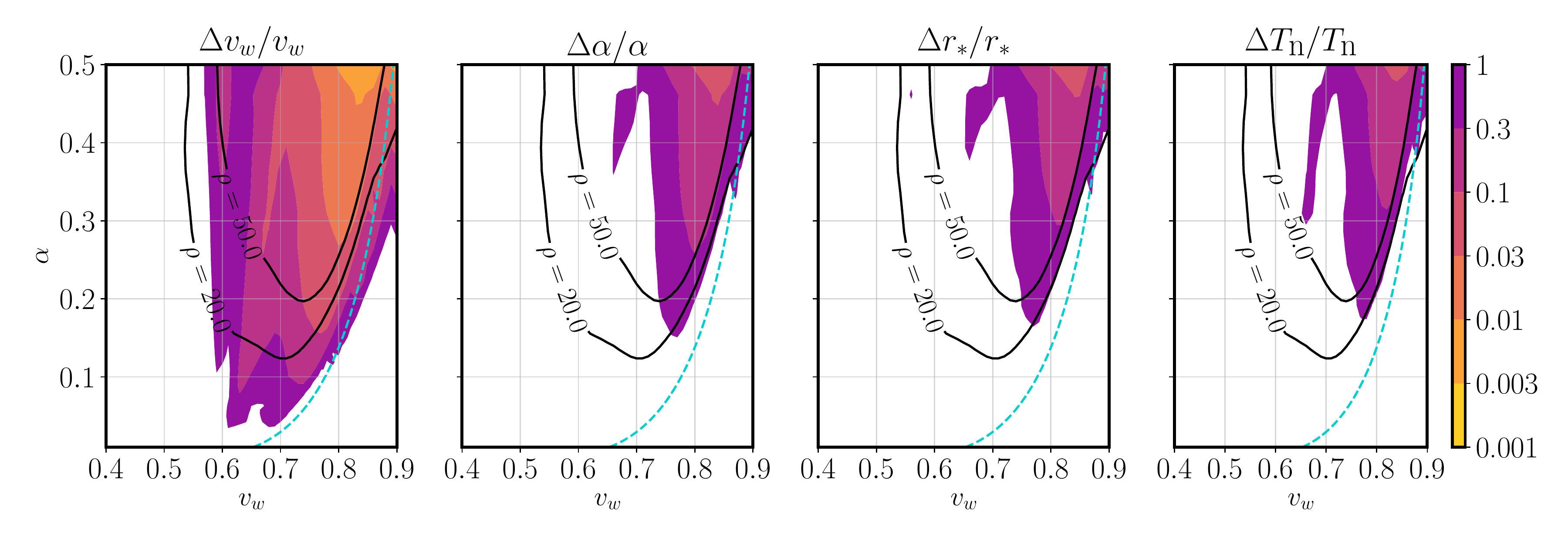}
\caption{\label{fig:thermo_params_fixed_rs_LISA_BBH_GB_T_100_as_param} Noise model: LISA instrument noise, foregrounds from extragalactic compact binaries Eq.~(\ref{Eq:Om_LV}) and unresolved galactic compact binaries (\ref{Eq:Om_gb}).
}
\end{subfigure}

\begin{subfigure}[b]{\textwidth}
\centering 
\rotatebox{90}{\hspace{20mm}$r_* = 0.01$}
\includegraphics[width=0.8\textwidth]{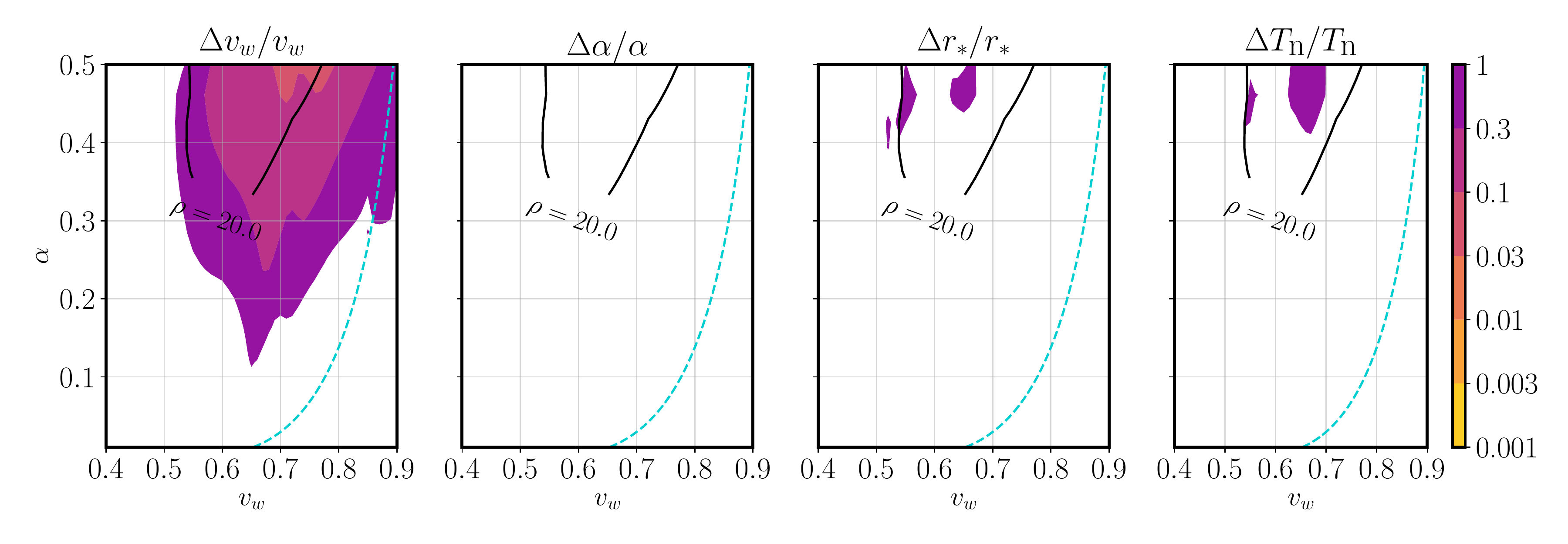}\\[-10pt]
\rotatebox{90}{\hspace{20mm}$r_* = 0.1$}
\includegraphics[width=0.8\textwidth]{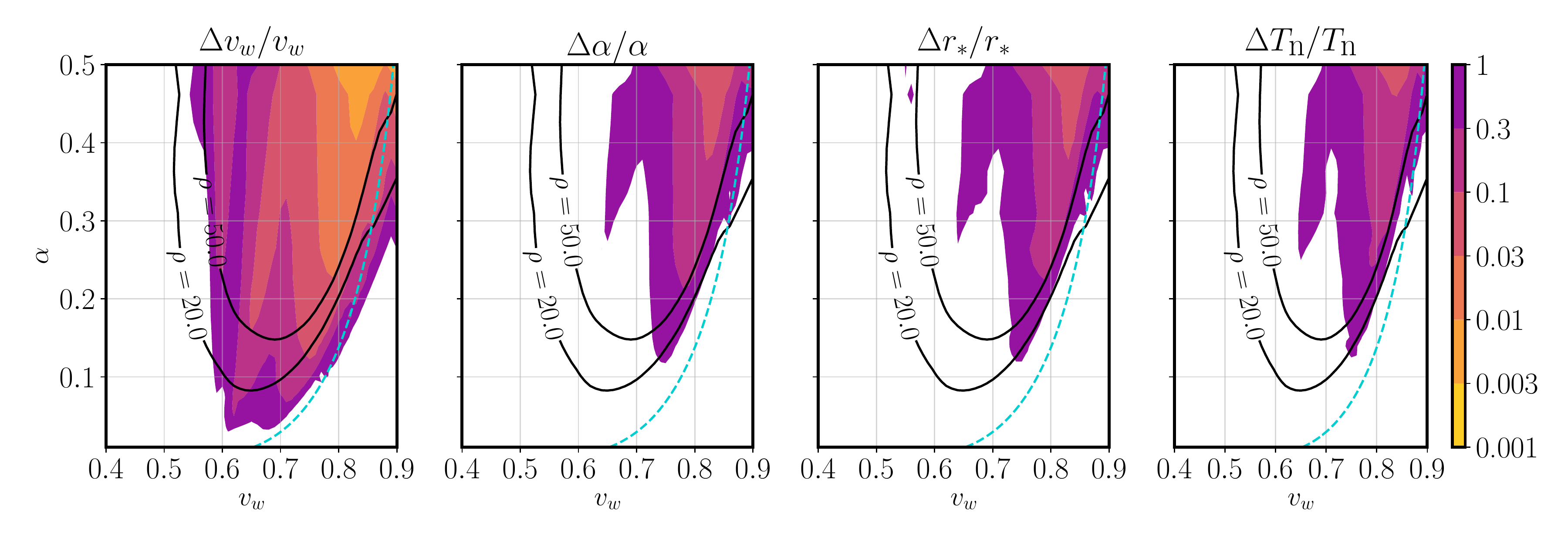}
\caption{\label{fig:thermo_params_fixed_rs_LISA_BBH_T_100_as_param} Noise model: same as above, 
with the foreground from unresolved galactic binaries removed.}
\end{subfigure}

\caption{\label{fig:thermo_params_fixed_rs_LISA_BBH_GB_T 100_as_param } Contours of relative uncertainty in the thermodynamic parameters wall speed $\vw$, transition strength $\al$, scaled mean bubble spacing $r_*$ and nucleation temperature $\TN$. 
In each sub-figure, the upper and lower panels have 
Hubble-scaled bubble spacing  $r_*$ as annotated. In both panels $\TN= 100$GeV. 
The black solid line shows contours of signal-to-noise ratio $\SNR$. 
The turquoise dashed line is the Jouguet speed, the minimum for a detonation. }
\end{figure}

The resulting relative uncertainties are shown 
 in Figs.~\ref{fig:thermo_params_fixed_rs_LISA_BBH_GB_T_100_as_param} with the galactic binary foreground and \ref{fig:thermo_params_fixed_rs_LISA_BBH_T_100_as_param} without the galactic binary foreground. Below the Jouguet speed, indicated by a dashed line, 
the fluid shell becomes a supersonic deflagration, with a significant change in 
the sound wave power spectrum, and hence the gravitational wave power spectrum \cite{Hindmarsh:2019phv}.
Thus one expects to see features in the signal-to-noise ratio and the relative uncertainties 
to the left of this line. 
The intricate shape of the contours is also partly due to the complex degeneracies, discussed below, 
and inaccuracies in the interpolation of the numerically-determined GW suppression factor.

A general conclusion is that, even when $\rho=20$, the only parameter 
which has relative uncertainty less than 1 is the wall speed. 
That the wall speed $\vw$ is the best determined parameter is perhaps 
surprising, but it can be understood as follows.

Looking at the upper left panel of Fig.~\ref{Fig:parameter_vary}, one can see  
that varying the wall speed significantly changes the shape of the power spectrum, 
which none of the other parameters do. 
On the other hand, the other parameters have complex degeneracies. 
For example, $r_*$ and $\TN$ both affect the overall frequency scale, 
and $\al$ and $r_*$ both affect the overall amplitude of the power spectrum. 
Increasing $\TN$ (see Fig.~\ref{Fig:parameter_vary}, bottom right panel) 
shifts the peak frequency, which can be compensated by a combination 
of increasing $r_*$ (Fig.~\ref{Fig:parameter_vary}, bottom left panel) 
and reducing $\al$. 

Another general conclusion, clear from the comparison between Figs.~\ref{fig:thermo_params_fixed_rs_LISA_BBH_GB_T_100_as_param} and \ref{fig:thermo_params_fixed_rs_LISA_BBH_T_100_as_param}, is the importance of 
removing the galactic binary foreground for parameter estimation. 
The two figures represent the extremes of what can be achieved in practice.
The study of Ref.~\cite{Adams:2013qma} indicates that the annual variation of the 
galactic binary foreground will enable its near-complete removal, and 
so Fig.~\ref{fig:thermo_params_fixed_rs_LISA_BBH_T_100_as_param} 
is likely to be a better approximation. 

\FloatBarrier
\subsection{Principal component analysis }
\begin{figure}[h!]
\begin{subfigure}[b]{\textwidth}
\centering 
\rotatebox{90}{\hspace{20mm}$r_* = 0.01$}
\includegraphics[width=0.8\textwidth]{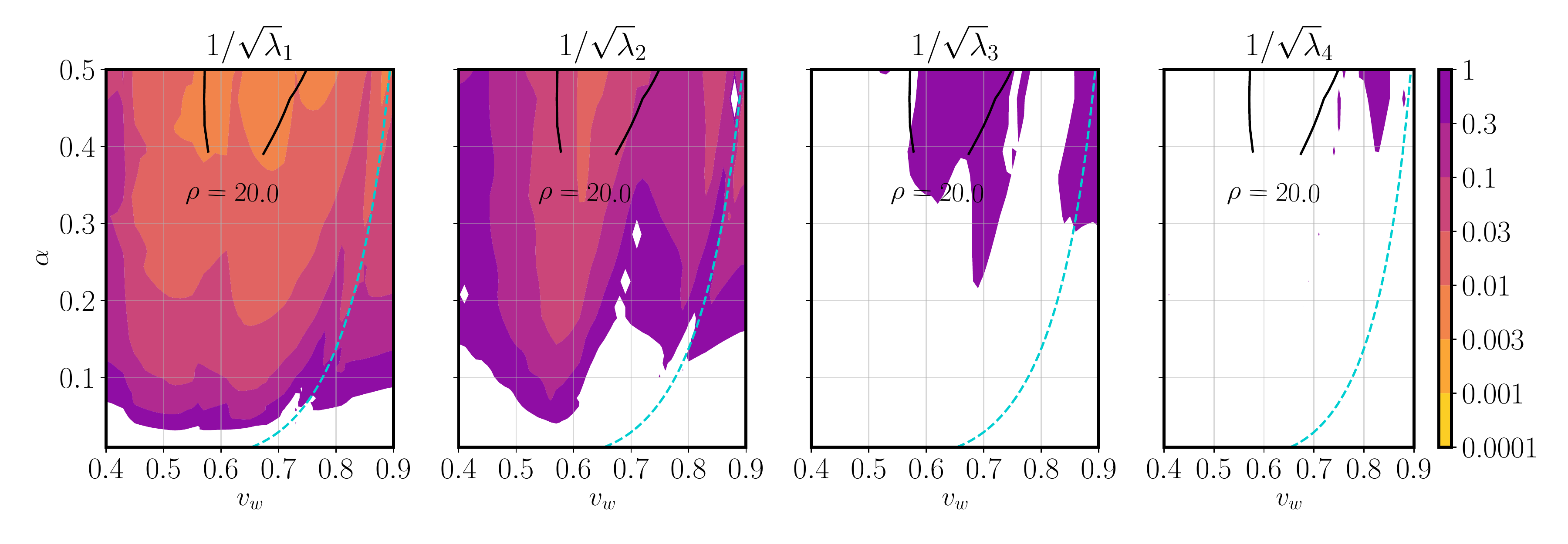}\\[-10pt]
\rotatebox{90}{\hspace{20mm}$r_* = 0.1$}
\includegraphics[width=0.8\textwidth]{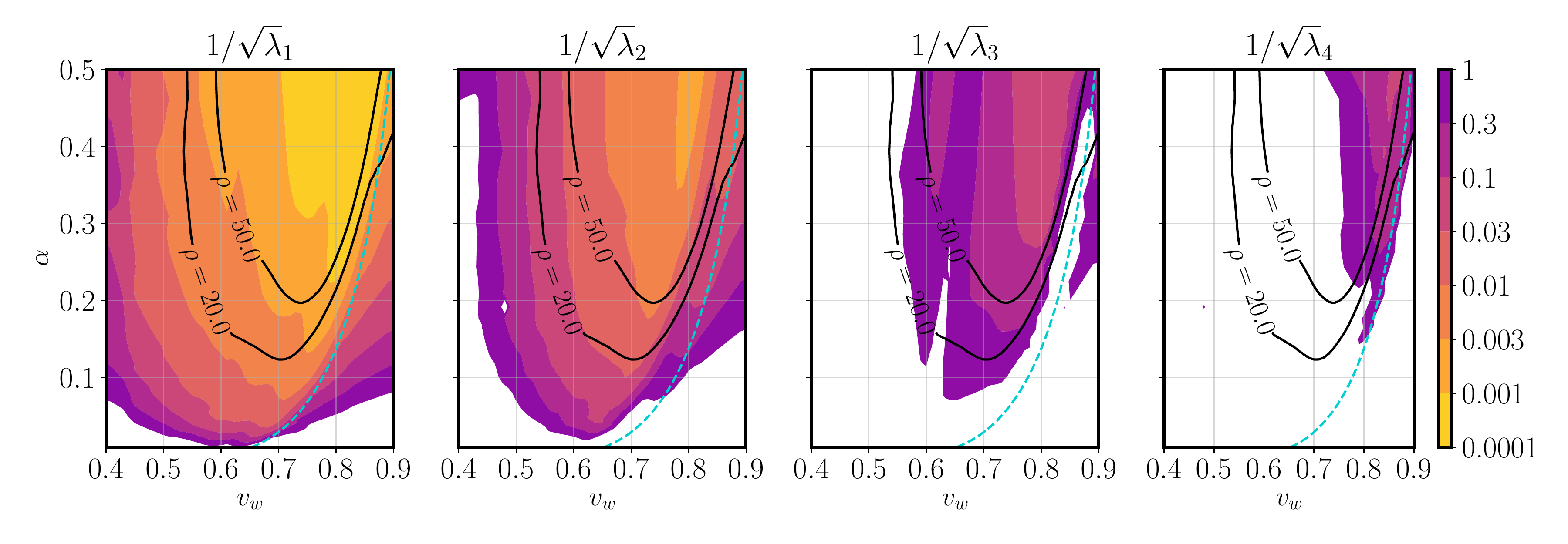}
\caption{\label{fig:eigenalues_fixed_rs_LISA_BBH_GB_T_100_as_param} Noise model: LISA instrument noise, foregrounds from extragalactic compact binaries Eq.~(\ref{Eq:Om_LV}) and unresolved galactic compact binaries (\ref{Eq:Om_gb}).
}
\end{subfigure}

\caption{\label{fig:principal_component_uncertainty_fixed_rs_LISA_BBH_GB_T 100_as_param} Contours of standard deviation ($1/\sqrt{\la_n}$) for the principal components constructed from the eigenvectors of the Fisher matrix evaluated across the wall speed $\vw$ and  phase transition strength $\al$ parameter space.
In each sub-figure, the upper and lower panels have Hubble-scaled bubble spacing  $r_*$ as annotated. In both panels $\TN= 100$ GeV. 
The black solid line shows contours of signal-to-noise ratio $\SNR$. 
The turquoise dashed line is the Jouguet speed, the minimum for a detonation. }

\end{figure}

The degeneracy between $\al$, $r_*$ and $\TN$ gives the impression that they will be virtually no sensitivity to these parameters, even at high signal-to-noise ratio. The Fisher matrix may be overestimating the uncertainties, so we look to the principal components  to see if there is greater sensitivity to linear combinations of the thermodynamic parameters.  The contours of the standard deviation of our principal components $\la_n^{-1/2}$ can be seen in Fig.~\ref{fig:principal_component_uncertainty_fixed_rs_LISA_BBH_GB_T 100_as_param}

Comparing Figs.~\ref{fig:thermo_params_fixed_rs_LISA_BBH_GB_T_100_as_param} and ~\ref{fig:principal_component_uncertainty_fixed_rs_LISA_BBH_GB_T 100_as_param} it is immediately obvious that there is greater sensitivity to the principal components over a broader region of parameter space, even when the foreground from galactic binaries is present.  In general, for GW power spectra with $\SNR > 20$ the two highest-order principal components reach $1/\sqrt{\la_n} < 3\% $ for both values of the Hubble-scaled mean bubble spacing. For GW power spectra with $r_* = 0.01$, 
whilst there is only a small region of sensitivity to $\vw$, there is broad sensitivity to the two highest-order principal components. 
In the $r_* = 0.1$ case $1/\sqrt{\la_n }< 30\%$ for the majority of the parameter space for the two highest-order principal components (see Fig.~\ref{fig:principal_component_uncertainty_fixed_rs_LISA_BBH_GB_T 100_as_param}).
\FloatBarrier
To investigate the contribution of the principal components to the thermodynamic parameters,  
we assigned to each of the first three principal components the colours red, green and blue respectively.
We took the four thermodynamic parameter eigenvectors in the principal component basis, 
and constructed an RGB colour from the square of the corresponding entry in the eigenvector. 
A significant mixture of the fourth principal component would then appear as a dark colour.

\begin{figure}[h!]

\centering 
\rotatebox{90}{\hspace{20mm}$r_* = 0.01$}
\includegraphics[width=0.8\textwidth]{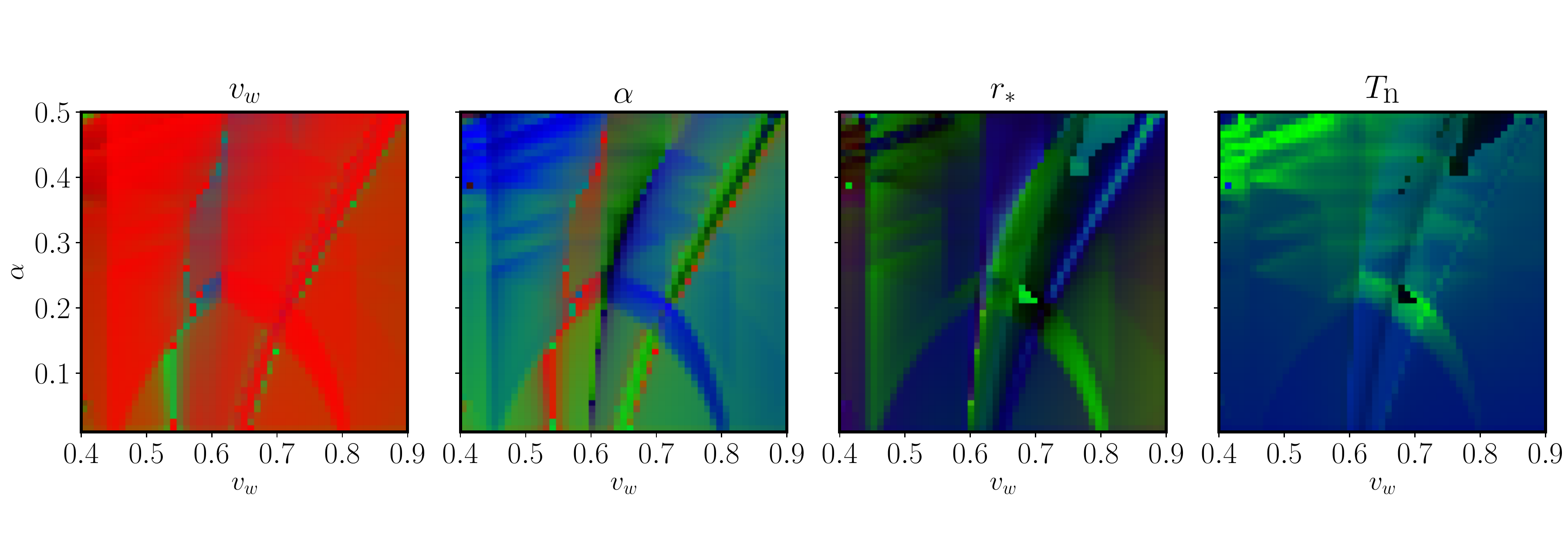}\\[-10pt]
\rotatebox{90}{\hspace{20mm}$r_* = 0.1$}
\includegraphics[width=0.8\textwidth]{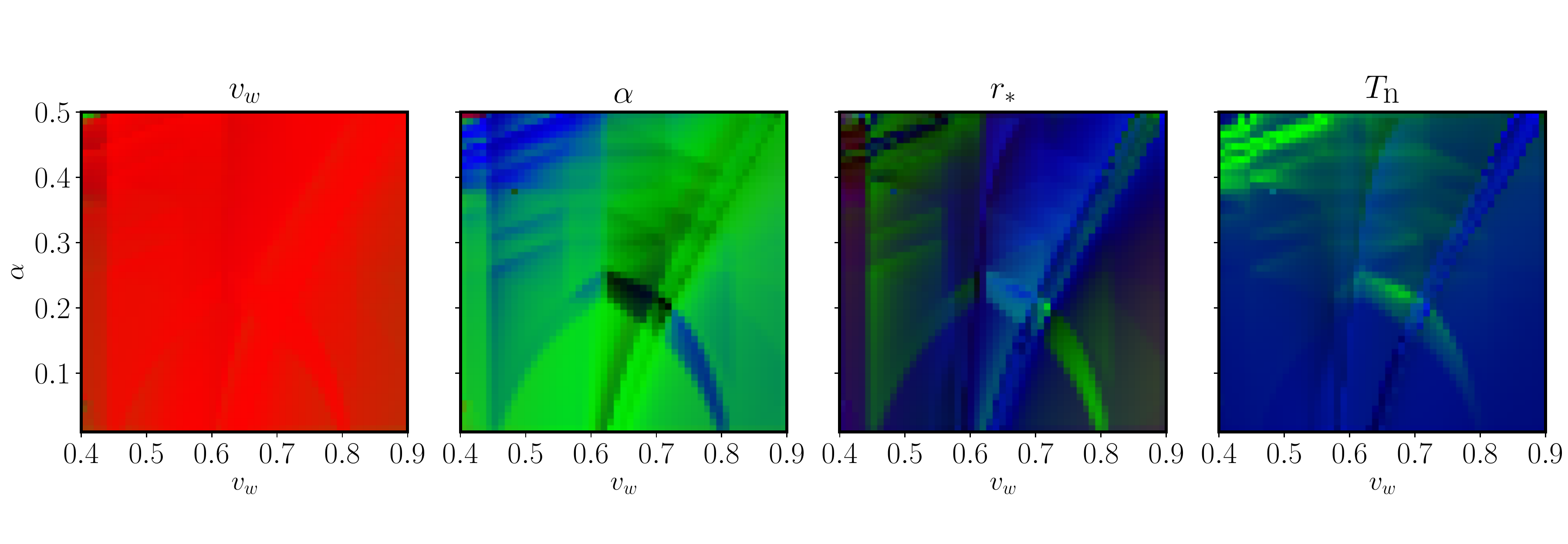}

\caption{\label{fig:eigen_vector} The contributions of the first three principal components to the thermodynamic parameters wall speed $\vw$, transition strength $\al$, scaled mean bubble spacing $r_*$ and nucleation temperature $\TN$. Red, green and blue correspond to the first, second and third principal components respectively.  The upper and lower panels have Hubble-scaled bubble spacing  $r_*$ as annotated. In both panels $\TN= 100$ GeV. Noise model: LISA instrument noise, foregrounds from extragalactic compact binaries Eq.~(\ref{Eq:Om_LV}) and unresolved galactic compact binaries (\ref{Eq:Om_gb}).
}
\end{figure}

We show the result in  Fig.~\ref{fig:eigen_vector}. 
We see the wall speed $\vw$ is predominantly red, meaning the first principal component provides the largest contribution, which confirms that we would expect greatest sensitivity to $\vw$.
The other parameters show an interesting mix of colours, which is partly noise introduced when we interpolate the kinetic energy suppression data (see Appendix \ref{sec:suppression_factor}). We believe the remaining sudden changes of colour comes from the degeneracy between parameters, in particular the streak originating around the speed of sound on the wall speed axis.   
\FloatBarrier

\subsection{Sound shell model with fixed nucleation temperature}
\label{ssec:TP_fixedT}

\begin{figure}[ht!]
\begin{subfigure}[b]{\textwidth}
\centering 
\rotatebox{90}{\hspace{20mm}$r_* = 0.01$}
\includegraphics[width=0.8\textwidth]{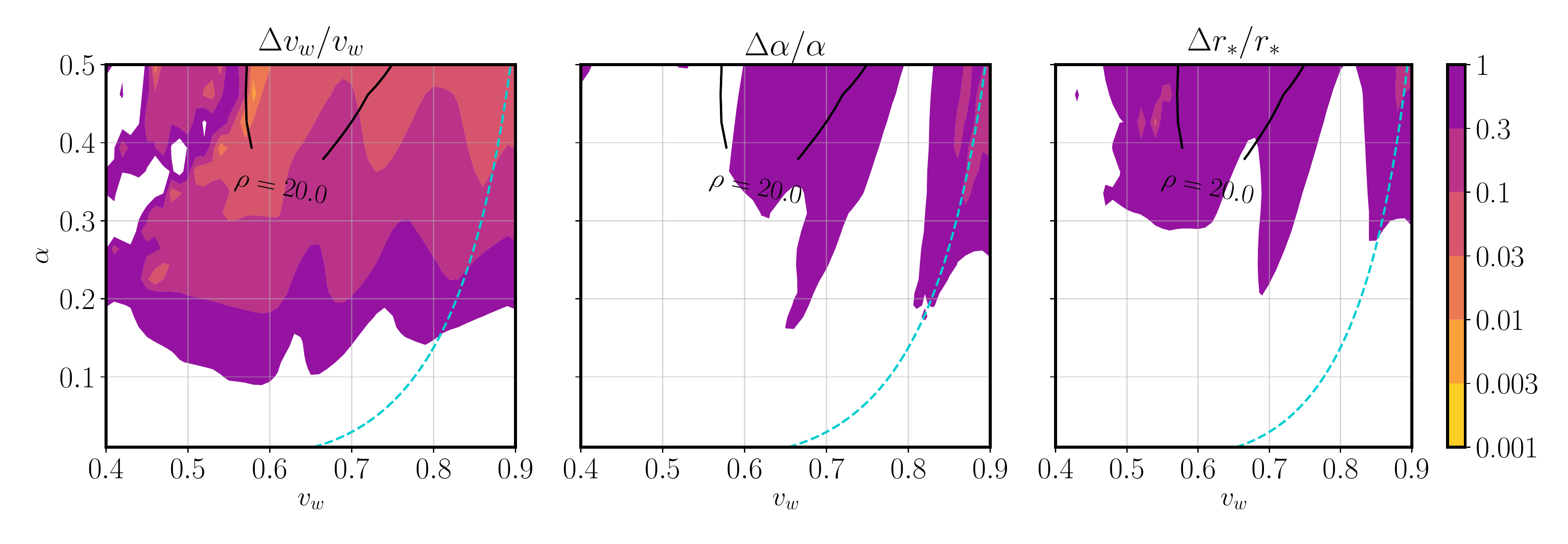}\\
\rotatebox{90}{\hspace{20mm}$r_* = 0.1$}
\includegraphics[width=0.8\textwidth]{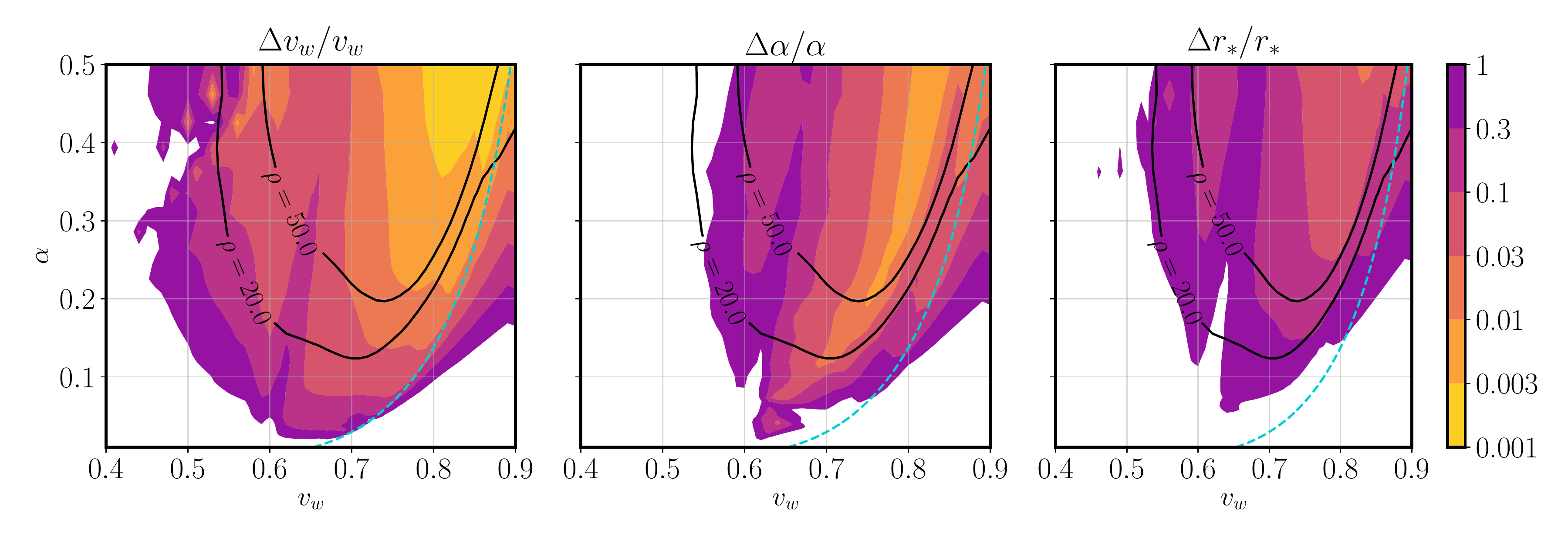}
\caption{
Noise model: LISA instrument noise, foregrounds from extragalactic compact binaries Eq.~(\ref{Eq:Om_LV}) and unresolved galactic compact binaries (\ref{Eq:Om_gb}).
}
\label{fig:thermo_params_fixed_T_BBH_GB_KE_suppress} 
\end{subfigure}
\begin{subfigure}[b]{\textwidth}
\centering 
\rotatebox{90}{\hspace{20mm}$r_* = 0.01$}
\includegraphics[width=0.8\textwidth]{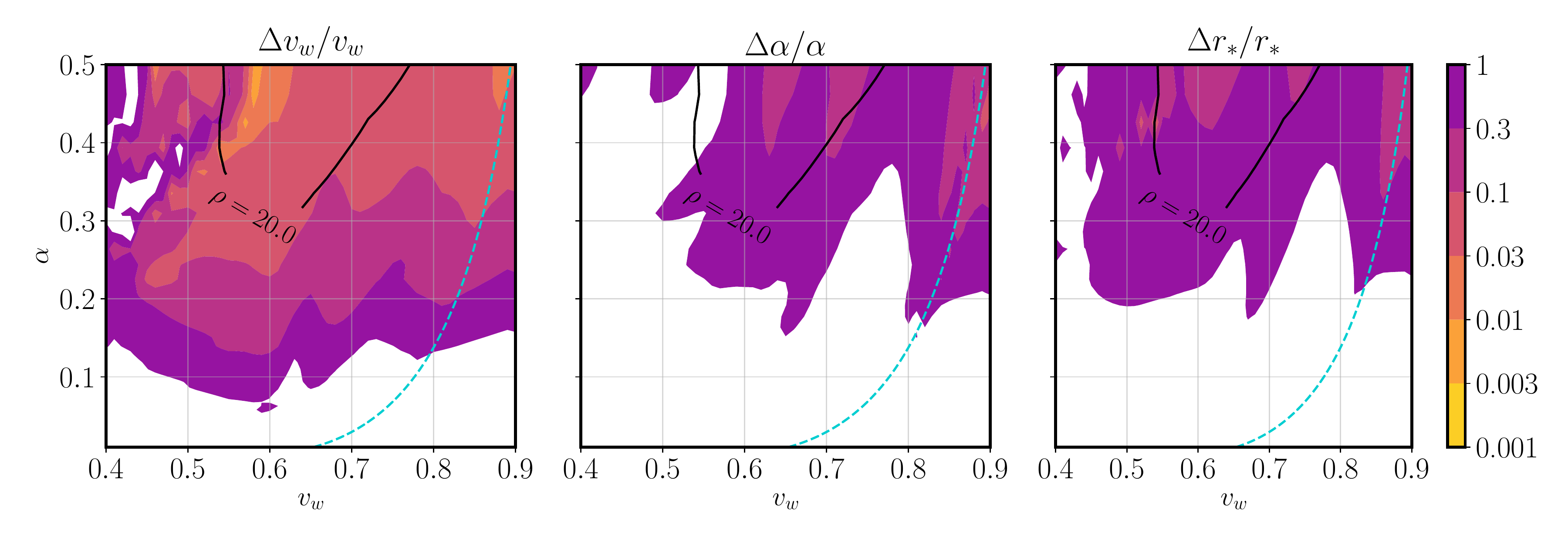}\\
\rotatebox{90}{\hspace{20mm}$r_* = 0.1$}
\includegraphics[width=0.8\textwidth]{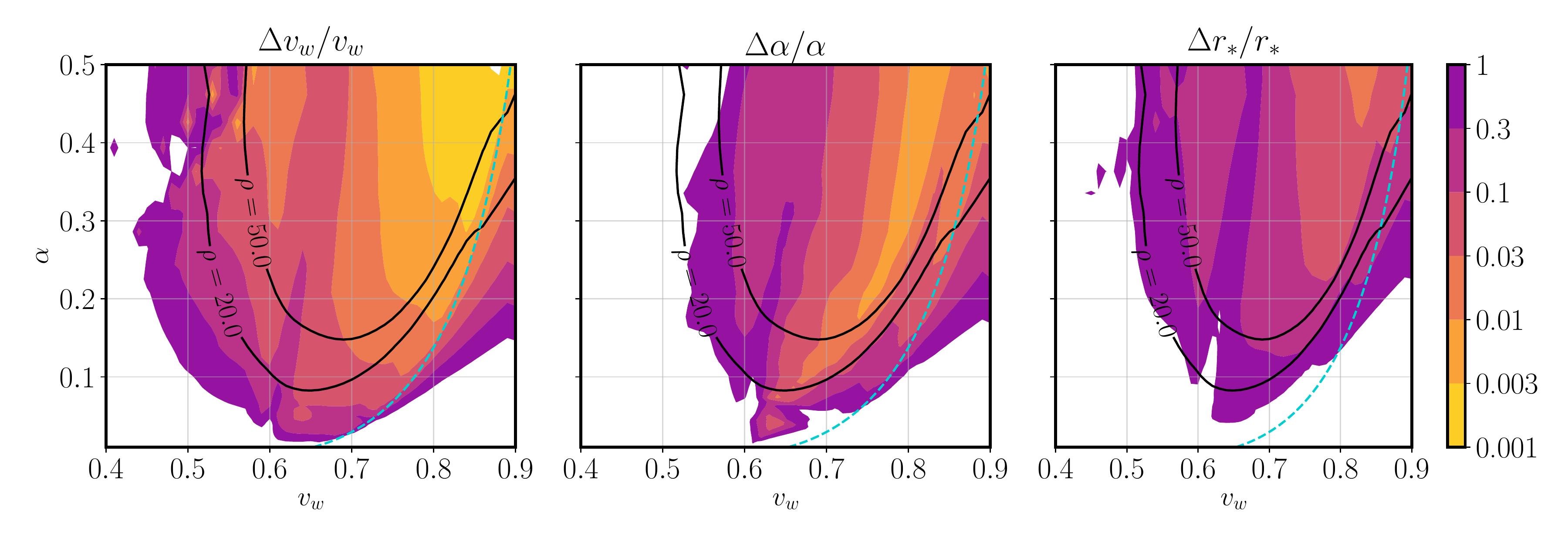}
\caption{Noise model same as above, 
with the foreground from unresolved galactic binaries removed.}
\label{fig:thermo_params_fixed_T_BBH_KE_suppress}
\end{subfigure}
\caption{Contours of relative uncertainty in the thermodynamic parameters wall speed $\vw$, transition strength $\al$ and scaled mean bubble spacing $r_*$ with 
nucleation temperature $\TN= 100$ GeV, for gravitational wave power spectra calculated using the sound shell model, Eq.~(\ref{Eq:Omgw0_sup}). 
In each sub-figure, 
the upper and lower panels have Hubble-scaled bubble spacing $r_*$  as annotated. The turquoise dashed line is the Jouguet speed, the minimum for a detonation.
}
\end{figure}

For the final analysis we explore the impact of information from particle physics data. 
While the information is likely to constrain a combination of parameters, we 
take as a limiting example a known nucleation temperature $\TN = 100$ GeV. 
The nucleation temperature is likely to be close to the critical temperature of the phase transition, 
which is the most straightforward thermodynamic parameter to calculate from an underlying theory,

The Fisher matrix, covariance matrix and relative uncertainties are calculated following the same procedure as above for two scenarios: first with our base noise model Fig.~\ref{fig:thermo_params_fixed_T_BBH_GB_KE_suppress} and then with the unresolved galactic binaries foreground removed, Fig.~\ref{fig:thermo_params_fixed_T_BBH_KE_suppress}.

\FloatBarrier

Prior knowledge of the nucleation temperature $\TN$ greatly improves the power of LISA to estimate other parameters, as the degeneracies become partly broken. With fixed $\TN$ for GW power spectra with $\SNR>20$ the wall speed has relative uncertainty of less than $10 \%$. Fixed $\TN$ also improves sensitivity to the phase transition strength, $\al$, and the  Hubble-scaled mean bubble spacing, $r_*$.
For example, if the phase transition has $r_* = 0.1$, one can achieve relative uncertainties  $\Delta\al/\al <10\%$ and $\Delta r_*/r_* <3 0\%$ 
with a signal-to-noise ration greater than $50$.

There is an interesting feature in the relative uncertainty contours at $r_* = 0.1$, where the SNR is higher:
a small ridge of lower uncertainty in $\alpha$, 
for wall speeds just over 0.6. 
This is accompanied by reduction in the sensitivity to $\vw$.  

The origin of this ridge is perhaps as follows. 
Referring to Fig.~\ref{Fig:parameter_vary}, one can see that 
at around $\vw = 0.6$ at $r_* = 0.1$, changes in the wall speed and $r_*$ 
have the effect of moving the closest part of the signal to the sensitivity curve 
in a direction tangent to the sensitivity curve, without changing the shape.  
This would mean that the likelihood changes little in these directions, and it would be difficult to distinguish between possible parameter values, this would lead to a reduction in sensitivity.  
Changes in $\al$, on the other hand, change the signal power, and 
will change the likelihood.  Thus the likelihood is most sensitive 
to changes in $\al$ in this region.

\section{Discussion}
\label{sec:Discussion}
In this paper we have explored the prospect of extracting the model parameters of a stochastic gravitational wave background from a first order phase transition at future space-based gravitational wave observatories. We focused on LISA, 
and the impact of including expected foregrounds from compact binaries. 
Here we studied the gravitational wave power spectra predicted by the  sound shell model (SSM), and 
used Fisher matrix analysis to investigate the sensitivity both to the four parameters of a double broken power law approximation, 
and to the underlying thermodynamic parameters in the SSM. The key thermodynamic parameters are the nucleation temperature
$\TN$, the transition strength $\al$, the mean bubble spacing in Hubble units $r_*$ and the wall (phase boundary) speed $\vw$. 
We assumed a sound speed $\cs =1/\sqrt{3}$ in both phases, and leave an investigation of sensitivity to this parameter to 
future work.  The fact that different sound speeds significantly change the kinetic energy fraction of the fluid \cite{Giese:2020znk,Giese:2020rtr}
suggests that there will be sensitivity to this parameter as well.

In Sec.~\ref{ssec:cosmo_dbl_brkn} we studied the double broken power law approximation to the gravitational wave power spectrum, which was advocated in Ref.~\cite{Hindmarsh:2019phv}.  
It has parameters characterising the peak $\OmPeak$ and two frequency scales, the peak frequency $\fp$, and a lower ``break'' frequency $\fb = \rb\fp$.  
In its original form, the indices of the three 
power laws were fixed by arguments based on the limits of certain integrals. 
We introduced a fourth parameter, the spectral slope of the intermediate power law $b$, 
to improve the fit between $\fb$ and $\fp$ for phase transitions proceeding by supersonic deflagrations.
This form, given in Eq.~(\ref{Eq:omgw_dbl_brkn}), is a significant improvement on the single broken power law in fitting the predictions of the 
sound shell model (see Fig.~\ref{fig:mean_res}).  

We performed a Fisher matrix analysis to calculate the relative uncertainty for the four parameters of the 
double broken power law spectrum. 
In Fig.~\ref{fig:obs_P_rel_uncer} we see that the  $\OmPeak$ and the peak frequency $\fp$  are expected to be best constrained, 
with a signal-to-noise ratio of 20 delivering determinations to around 10\% 
for peak frequencies between $10^{-4}$ and $10^{-2}$ Hz. 
The other parameters, the break frequency ratio $\rb$ and the intermediate power law $b$, are less well determined, but 
can be determined with less than 1\% relative uncertainty for signals with peak power and peak frequencies that lie on 
LISA's sensitivity curve, that is, signals at the same level as or above the instrument noise.

The extragalactic compact binary foreground expected from LIGO/Virgo data \cite{Abbott:2017xzg} is not an important contributor to the total noise, 
but the galactic binary foreground would be significant if it could not be removed.
The main effect is to somewhat reduce the range over which parameters can be determined within a given uncertainty; 
the magnitude of the effect can be judged in the difference between Figs.~\ref{fig:thermo_params_fixed_rs_LISA_BBH_GB_T_100_as_param} and \ref{fig:thermo_params_fixed_rs_LISA_BBH_T_100_as_param}.
However, it is expected that the galactic binary foreground will be at least partially removable through its annual modulation \cite{Adams:2013qma}. 

We also studied LISA's sensitivity to the four principal thermodynamic parameters of a first order phase transition, as described above.  The GW power spectrum model used was the sound shell model, Eq.~(\ref{Eq:Omgw0_sup}),  incorporating  kinetic energy suppression in slow deflagrations \cite{Cutting:2019zws}. 

We investigated scenarios with a nucleation temperature $\TN = 100$ GeV and 
Hubble-scaled mean bubble spacing $r_* = 0.1, 0.01$, scanning over 
a range in $(\vw,\al)$ space with $0.4 \le \vw \le 0.9$, $0.005 \le \al \le 0.5$, where 
numerical simulations have been performed \cite{Cutting:2019zws} and the model can be calibrated. 
We observed that in order to match the total power in the simulations a suppression factor had to be applied 
to the sound shell model (see Appendix \ref{sec:suppression_factor}).
 
We found that the wall speed $\vw$ would be the best constrained parameter, with a relative uncertainty of better 
than $30$\% provided the signal-to-noise ratio is above 20, and the wall is supersonic, 
even in the worst-case scenario where the foreground from unresolved galactic binaries cannot be removed.
As the Hubble-scaled mean bubble spacing $r_*$ gets smaller the signal power decreases, this leads to a reduction in the region of parameter space over which a signal-to-noise ratio of 20 can be achieved.
For example, with $r_* = 0.1$, 
SNR 20 can be produced by phase transition strengths down to about $\al \simeq 0.13$, 
while at $r_* = 0.01$ the corresponding figure is $\al \simeq 0.35$.

There is limited sensitivity to the parameters $\al$, $r_*$ and $\TN$ due to 
degeneracies.  For example, the peak frequency is left unchanged by simultaneous changes in the
nucleation temperature $\TN$ and the mean bubble spacing $r_*$.  Changing $r_*$ changes the peak 
power, which can be brought back to its original value, without changing the peak frequency much, by a change in 
the transition strength $\al$. 
This will mean that the parameters most easily computable from underlying models, 
$\TN$, $\tilde\be$, and $\al$, will not be individually well determined.

However, there is much better sensitivity to the principal components 
across the explored parameter space. The two highest-order principal components have relative uncertainty  less than $3\%$ for GW power spectra with $\SNR> 20$, for both values of the Hubble-scaled mean bubble spacing.  The highest-order component is found to be dominated by the wall speed, 
as is consistent with the wall speed being the best determined parameter.

If one of the parameters is known, the other thermodynamic parameters are much better constrained. 
For example, with a known nucleation temperature $\TN = 100$ GeV, the wall speed would have an estimated uncertainty of less than $10\%$ for the majority of the parameter space, 
and the phase transition strength would  
be almost as accurately measured as the wall speed. 
The mean bubble spacing would be less accurately measured.  
A more realistic situation would involve constraints on masses and coupling constants in an underlying 
particle physics model, which we will explore elsewhere. 

The parameter degeneracies we have found mean that 
 one must consider the reliability of the Fisher matrix as an indicator of parameter uncertainties. 
 For the spectral parameters of the double broken power law fit the Fisher matrix can be trusted, as there is little or no degeneracy between parameters and the Gaussian approximation made with the Fisher matrix is reasonable. 
To check, we carried out preliminary Markov Chain Monte Carlo (MCMC) estimation using the Gaussian 
approximation to the likelihood, which returned ellipsoidal posteriors for the spectral parameters. We also found
approximate agreement between the marginalised posteriors and the uncertainties predicted by Fisher analysis. On the other hand the thermodynamic parameters do have significant degeneracies. In some regions of parameter space where the signal to noise ratio is high, 
 preliminary MCMC has shown ellipsoidal posteriors, supporting the case that the principal component analysis 
 of the Fisher matrix should be a reasonably good indicator.
The general tendency of the Fisher matrix is to over-estimate uncertainties \cite{Efstathiou:1998xx}, and so we expect 
that the uncertainty estimates presented here are conservative. 
  
In summary, we have presented 
the relative uncertainties calculated with the Fisher matrix as a preliminary guide to the expected power of LISA to resolve the spectral and thermodynamic parameters. The analysis makes it clear there are significant degeneracies that limit the accuracy of the direct 
determination of the thermodynamic parameters. However, the principal components show that at least two combinations 
of parameters can be well-determined, and the wall speed is will be the best measured phase transition parameter.  
For GW signals with signal-to-noise ratio greater than 20  we found the relative uncertainty for the two highest-order principal components to be less than $ 3\%$. This provides a target for the accuracy required from theoretical models.
We plan to carry out a more detailed MCMC analysis, exploring more realistic noise models, and with the sound speeds as parameters to an extended sound shell model.

\acknowledgments

We thank  Oliver Gould, Elina Keih\"anen, Antony Lewis, Jes\'us Torrado, Ville Vaskonen, Essi Vilhonen and Graham White for useful comments and feedback. We are grateful to Daniel Cutting for supplying data from Ref.~\cite{Cutting:2019zws} and F\"eanor Reuben Ares for helpful discussions. CG (ORCID ID 0000-0002-7955-4465) is supported by a STFC Studentship. MH (ORCID ID 0000-0002-9307-437X) acknowledges support from the Academy of Finland (grant number 333609).

\newpage

\appendix

\section{Kinetic energy suppression in the sound shell model}
\label{sec:suppression_factor}

Here, we outline the implementation of kinetic energy suppression in our model of the GW spectrum, which 
brings the sound shell model \cite{Hindmarsh:2019phv}  into better agreement with numerical 
simulations \cite{Cutting:2019zws} at low wall speeds and high 
transition strengths.

The gravitational wave power spectrum 
takes the form (\ref{Eq:Omgw_ssm}).  We can remove all dependence on the 
mean bubble separation and the nucleation temperature by considering the scale-free power spectrum  
\ben\label{e:PowSpeTil}
\PspecGWhat(z) \equiv \left(\HN \tau_{\mathrm{v}}\right)^{-1} \left(\HN R_*\right)^{-1} \OmGW(z)= 3K^{2}(\vw,\al)\frac{z^{3}}{2 \pi^{2}} \SpecDenGW\left(z\right) .
\een
Integrating this quantity over $\ln(z)$, we define a dimensionless gravitational wave production efficiency parameter $\OmGWscaled$
by dividing by square of the kinetic energy fraction around a single self-similar bubble, $K_1(\vw,\al)$, giving 
\ben {\label{Eq:om_tilde}}
 \int \frac{dz}{z} \, \PspecGWhat(z)  =  K^2_1(\vw,\al) \OmGWscaled .
\een
This integral can be compared between the sound shell model and the numerical simulations.  
We will assume that the numerical simulations give a better estimate of $ \OmGWscaled$ than the 
sound shell model, and so we scale the sound shell model power spectrum by the ratio 
of the production efficiency parameters.  

\begin{figure}[h!]
\centering 

\includegraphics[width=0.7\textwidth]{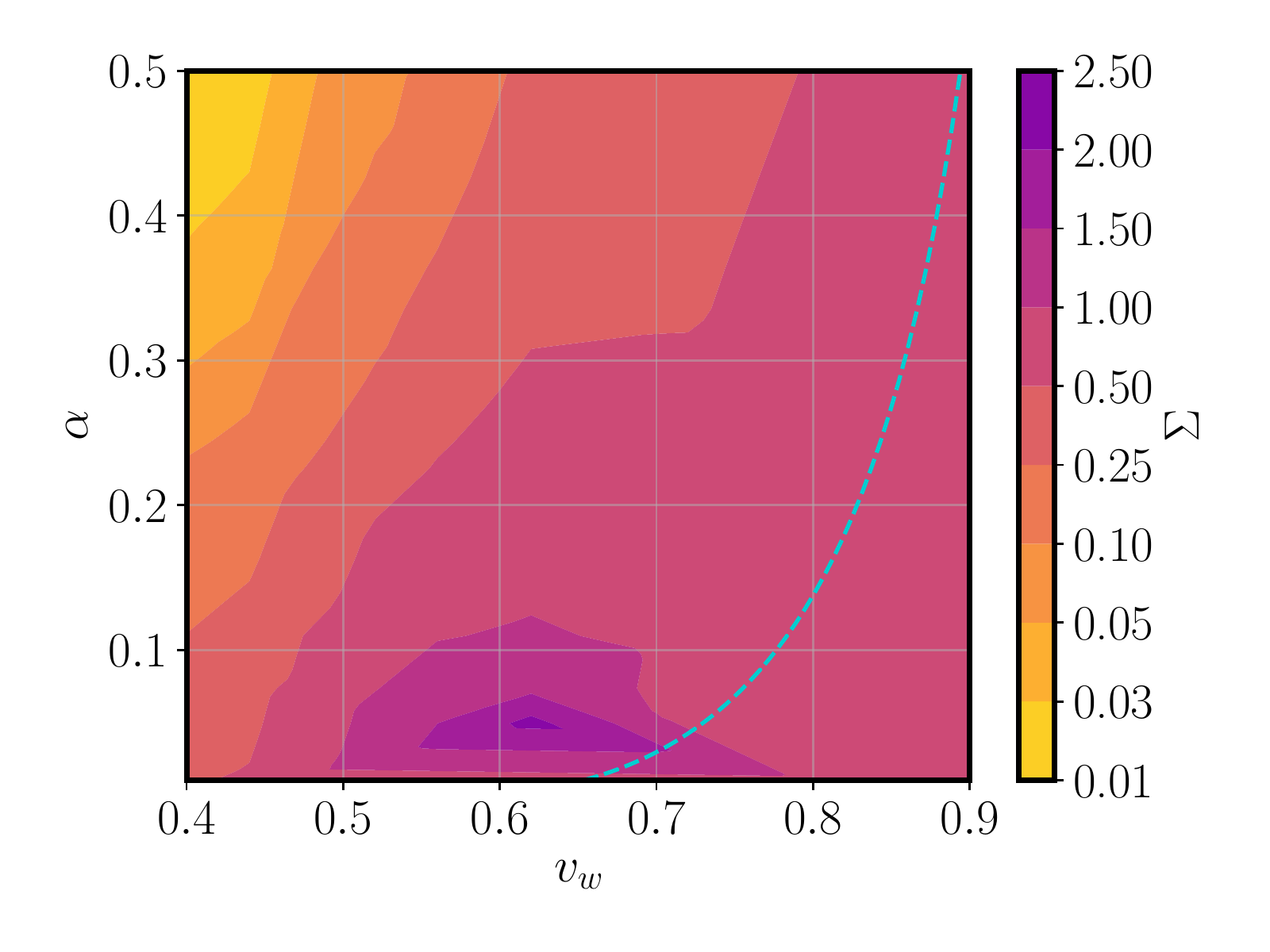}
\caption{\label{fig:sup_factor}  Contours showing the gravitational wave power suppression factor $\Sigma$
(\ref{e:SupFac}) in the $(\vw,\al)$ plane.  The suppression factor is constructed to make the 
total gravitational wave power in the sound shell model \cite{Hindmarsh:2019phv} agree with the simulations of Ref.~\cite{Cutting:2019zws}. The turquoise dashed line shows the Jouguet speed, the minimum speed for a detonation.} 
\end{figure}

A complication is that in the simulations, the bubbles have not yet reached self-similar profiles 
and their full kinetic energy fraction when 
they collide, and so the total gravitational wave power is underestimated.  
To compensate for this effect, 
the estimate of $\OmGWscaled^\text{sim}$ is made by dividing by the kinetic energy fraction around 
a numerical solution of the 1d hydro equations at a bubble size of $R_*^\text{sim}$, the mean bubble size found in the 
simulation.  We denote this kinetic energy fraction by $K_{1}(\vw,\al;R_*^\text{sim})$.  
It is related to the RMS fluid velocity $\bar{U}_\text{f,exp}$ given in the 7th column of Table 1 of 
Ref.~\cite{Cutting:2019zws} by
\ben
K_{1}(\vw,\al;R_*^\text{sim}) = \frac{4}{3} \bar{U}_\text{f,exp}^2.
\een
The suppression factor is then
\ben
\label{e:SupFac}
\Sigma(\vw,\al) = 
\frac{\int d\ln(z) \, \PspecGWhat^\text{sim}(z)}{\int d\ln(z) \, \PspecGWhat^\text{ssm}(z)}
\left( \frac{K_1(\vw,\al)}{K_{1}(\vw,\al;R_*^\text{sim})} \right)^2.
\een  
The simulations of Ref.~\cite{Cutting:2019zws} covered a region $0.24 \lesssim \vw \lesssim 0.92$ and $0.005 \lesssim \al \lesssim 0.5$ with around 60 samples, and gave $\int d\ln(z) \PspecGWhat^\text{sim}(z)$ in the second last column of Table 1. 
The resulting $\Sigma$ is plotted as coloured contours in the $(\vw,\al)$ plane in Fig.~\ref{fig:sup_factor}. 
Intermediate values are obtained by linear interpolation. 

\section{Broken power law approximations to the GW spectrum}
\label{sec:fit_compare}

\begin{figure}[h!]
\centering 
\includegraphics[width=0.45\textwidth]{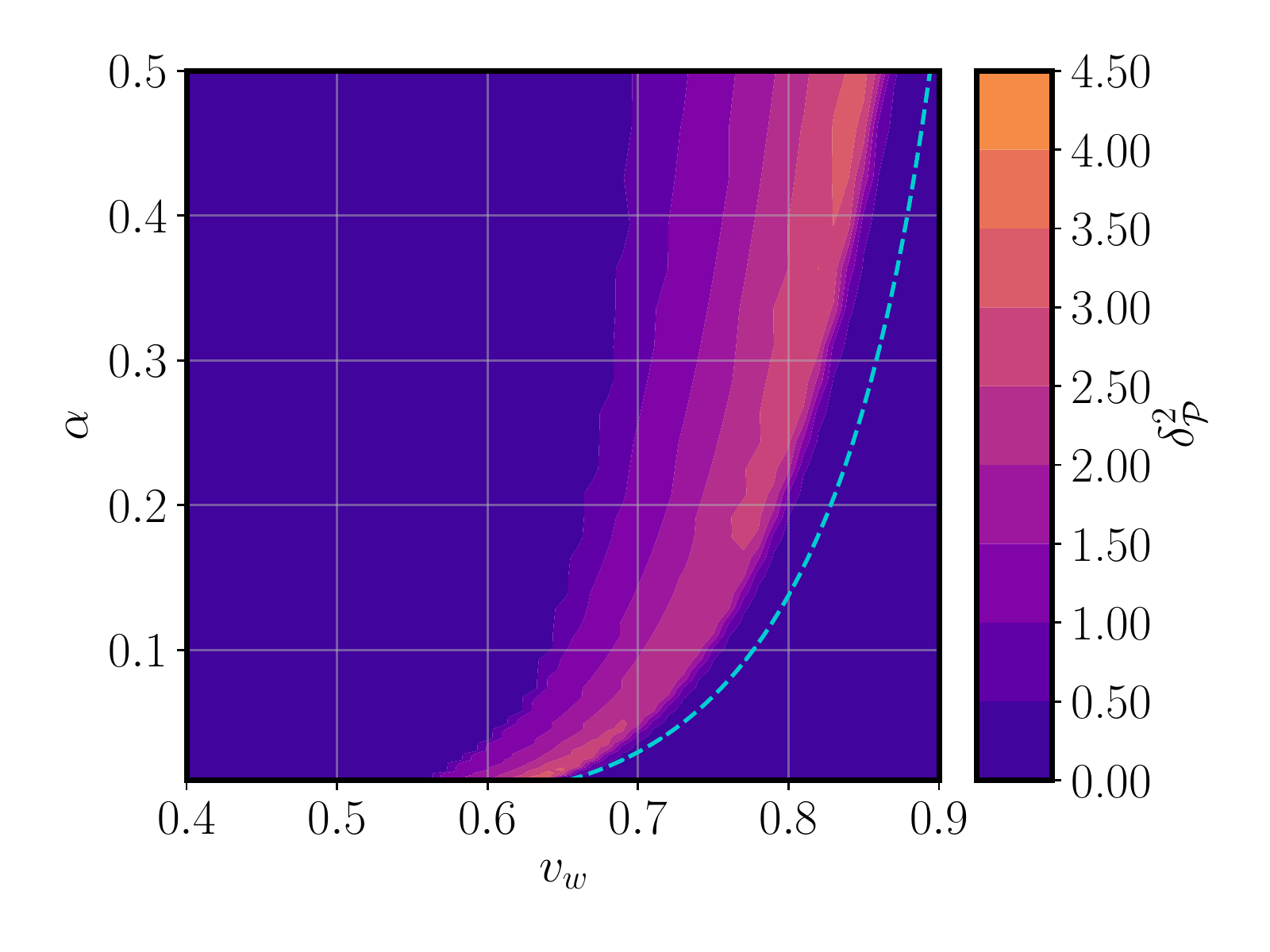}
\includegraphics[width=0.45\textwidth]{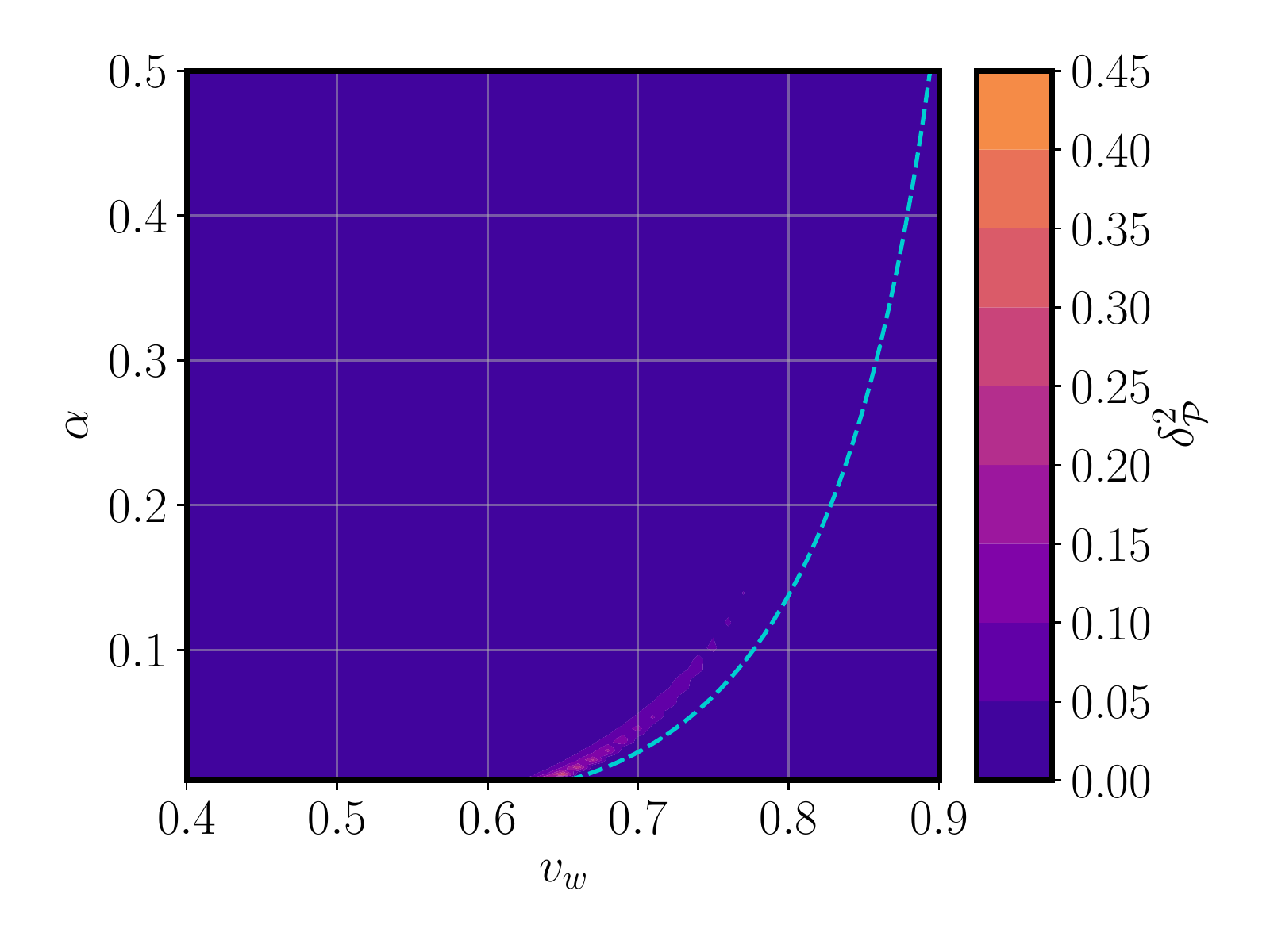}
\caption{\label{fig:mean_res} 
The mean squared relative deviation, $\de_\mathcal{P}^2$ defined in Eq.~(\ref{Eq:reduced_chi_squared}), for two analytic fits to the sound shell model gravitational wave power spectrum. On the left is the single broken power law used by the LISA cosmological working group \cite{Caprini:2019egz}, given in Eq.~(\ref{e:CWGBroPowLaw}).  On the right is the general double broken power law given in Eq.~(\ref{Eq: M double_break}). The turquoise dashed line is the Jouguet speed, Eq.~(\ref{Eq:Jouguet}), the minimum speed of the phase boundary in a detonation. Note the difference in the colour scales.  }
\end{figure}

Here we compare the single broken power law  
and the double broken power law approximations to the GW power spectrum from a first order phase transition as described by the sound shell model.  For each combination of thermodynamic parameters we find the best fit spectral parameters $\vec{\th}$ 
for each analytic form. We evaluate the minimum of the 
mean-squared relative deviation of the scale-free power spectra (see Eq.~\ref{e:PowSpeTil}),
\ben{\label{Eq:reduced_chi_squared}}
\de_\mathcal{P}^2  = \min_{\vec{\th}} 
\int \frac{dz}{z} \,  \left( \frac{\PspecGWhat^\text{fit}(z,\vec{\th}) - \PspecGWhat^\text{ssm}(z;\vw,\al)}{\PspecGWhat^\text{ssm}(z;\vw,\al)} \right)^{2},
\een
where $\PspecGWhat^\text{fit}$ is the power law fit to the scaled power spectrum calculated in the SSM $\PspecGWhat^\text{ssm}$, which depends only on $\vw$ and $\al$. 
We evaluate this quantity for the two different fit functions: 
 the four-parameter double broken power law Eq.~(\ref{Eq: M double_break}) and
the two-parameter broken power law 
given by the LISA Cosmology working group \cite{Caprini:2019egz}, where $M(s,\rb,b)$ is replaced by 
\ben
\label{e:CWGBroPowLaw}
C(s) = \frac{7 s^3}{(4 + 3 s^2)^{7/2}} . 
\een
The result is plotted in Fig.~\ref{fig:mean_res}. 

One sees that the two-parameter fit is poor for supersonic deflagrations, where the peak in the power spectrum is broad, 
while the four-parameter double broken 
power law has a mean square deviation less than $0.05$ almost everywhere.

\FloatBarrier

\bibliography{GW_EW_data_analysis}

\end{document}